\documentclass[12pt]{article}

\usepackage{amsmath,amssymb}
\usepackage{graphicx}
\usepackage{caption}
\usepackage{color}
\usepackage{dcolumn}
\usepackage{bm}
\usepackage[numbers,super,comma,sort&compress]{natbib}

\definecolor{background-color}{gray}{0.98}

\usepackage[titletoc,title]{appendix}
\usepackage{etoolbox}
\patchcmd{\appendices}{\quad}{: }{}{}
\usepackage{nameref}
\GetTitleStringSetup{expand}
\GetTitleStringDisableCommands{%
  \let\Large\empty
  \let\large\empty
  \let\sffamily\empty
  \let\color\@gobble 
}

\newcommand{\mysection}[1]{\section{\sffamily \Large #1}}
\newcommand{\mysubsection}[1]{\subsection{\sffamily \large #1}}
\newcommand{\mysubsubsection}[1]{\subsubsection{\sffamily \normalsize #1}}
\newcommand{\myref}[1]{Sec.~\ref{#1}}

\newcommand{\myfigs}[1]{#1}

\usepackage[draft,inline,nomargin,index]{fixme}
\fxsetup{theme=color,mode=multiuser}
\FXRegisterAuthor{dj}{andj}{DJ}
\FXRegisterAuthor{aw}{anaw}{AW}
\providecommand{\tabularnewline}{\\}
\usepackage[acronym]{glossaries}
\makeglossaries
\newacronym{ks}{KS}{Kohn-Sham}
\newacronym{diis}{DIIS}{direct inversion in the iterative subspace}
\newacronym{scf}{SCF}{self-consistent field}
\newacronym{tddft}{TDDFT}{time-dependent density functional theory}
\newacronym{dft}{DFT}{density functional theory}
\newacronym{qoct}{QOCT}{quantum optimal control theory}
\newacronym{tdcdft}{TDCDFT}{time-dependent current density functional theory}
\newacronym{ptddft}{P-TDDFT}{partition time-dependent density functional theory}
\newacronym{tn}{TN}{truncated-Newton}
\newacronym{oep}{OEP}{optimized effective potential}
\newacronym{pde}{PDE}{partial differential equation}
\newacronym{dcbc}{DCBC}{density-constrained boundary conditions}

\renewcommand\Re{\operatorname{Re}}

\usepackage{amsmath}
\usepackage{amsfonts}
\usepackage{bm}
\usepackage{esint}
\usepackage{braket}
\usepackage[linkcolor=black,pdfusetitle]{hyperref}

\title{Numerical Methods for the Inverse Problem of Density Functional Theory}
\author{Daniel S. Jensen\thanks{Center for Computing Research MS 1321, Sandia National Laboratories, Albuquerque, New Mexico, 87185-1321 E-mail: dsjense@sandia.gov} \,and Adam Wasserman\thanks{Department of Chemistry, Purdue University, West Lafayette, Indiana 47907, USA} \thanks{Department of Physics and Astronomy, Purdue University, West Lafayette, Indiana 47907, USA}}

\begin{document}

\maketitle


\begin{abstract}
The inverse problem of Kohn-Sham \gls{dft} is often solved in an effort to benchmark and design approximate exchange-correlation potentials.
The forward and inverse problems of \gls{dft} rely on the same equations but the numerical methods for solving each problem are substantially different.
We examine both problems in this tutorial with a special emphasis on the algorithms and error analysis needed for solving the inverse problem.
Two inversion methods based on \acrlong{pde} constrained optimization and constrained variational ideas are introduced.
We compare and contrast several different inversion methods applied to one-dimensional finite and periodic model systems.

\end{abstract}

\clearpage


  \makeatletter
  \renewcommand\@biblabel[1]{#1.}
  \makeatother

\bibliographystyle{apsrev}

\renewcommand{\baselinestretch}{1.5}
\normalsize

\clearpage

\glsresetall
\mysection{Introduction}\label{sec:introduction}
%
\Gls{dft} is a very popular and formally exact quantum many-body theory that has successfully been used to model both finite and periodic systems.\cite{martin_electronic_2004,burke_dft_2013}
Much of this success stems from the availability of approximate exchange-correlation functionals used as input in the direct problem of \gls{dft} to produce the ground-state electronic density of a given system.
Although far less common, this process can be reversed and the true exchange-correlation potential can be recovered from a given electronic density by solving the inverse problem of \gls{ks} \gls{dft}.\footnote{Other flavors of \gls{dft} exist but we focus on the standard ground-state version of \gls{ks} \gls{dft} and drop the \gls{ks} prefix for brevity throughout this tutorial.}
Although both the forward and inverse problems use the same equations, the difference in input and output variables leads to major differences in solution methods for the two problems.
We refer to the inverse problem of \gls{dft} as a density-to-potential inversion and the numerical methods required for solving this problem are the main focus of this tutorial.
In this section we give a brief introduction to inverse problems and explain how density-to-potential inversions can be used within \gls{dft}.

Inverse problems are common in science and have been central in quantum mechanics since its inception.
Much of what we know about the structure of matter has come from scattering experiments,\cite{merzbacher_quantum_1998} which can be described mathematically as inverse problems.
A variety of mathematical and numerical techniques exist for solving scattering problems as well as other inverse problems.\cite{kirsch_introduction_2011}
These methods are often very different from the methods used for solving the direct problems due to the differences in mathematical structure and input data.

According to Hadamard,\cite{hadamard_sur_1902} a problem is well-posed if a solution exists, it is unique, and it depends continuously on the data.
If any of the three properties listed above is violated then the problem is ill-posed.
The forward or direct problem of \gls{dft} is widely considered to be well-posed thanks to several proofs including the Hohenberg-Kohn Theorems and the constrained-search formulation of Levy and Lieb.\cite{martin_electronic_2004} 
The inverse problem of \gls{dft} can also be well-posed for discretized systems as proven in Ref.~\citenum{chayes_density_1985} but often errors and missing information in the input density lead to violations of Hadamard's well-posed conditions.
In such cases, special precautions must be taken in designing inversion methods that converge to the true solution of the inverse problem.

We have found that understanding the numerical errors involved in the inverse problem of \gls{dft} is extremely helpful in developing successful \gls{dft} inversion methods.
Appendix \ref{app:numerical-issues} and Sec.~\ref{sec:one-orb} contain a number of examples showing how the numerical error in both the inversion algorithm and the target density being inverted set bounds on the inverted potential's accuracy.
A central theme in this tutorial is the creation of robust inversion methods that use information about the numerical errors involved in order to produce accurate potentials.
This information can be used to properly smooth target densities or choose a form of regularization.
Regularization is the general term used to describe a method for solving ill-posed problems.
Some regularization methods are as simple as stopping an optimization according to an a posteriori parameter choice rule while others, such as total variation, can be sophisticated and very challenging mathematical and programming exercises.\cite{vogel_computational_2002}
Often the form of regularization is chosen to penalize features known to be incorrect based on a priori information about the solution.
For a thorough review of regularization and inverse problem theory see Refs.~\citenum{vogel_computational_2002,aster_parameter_2012,muller_linear_2012}.
The density-to-potential inversion methods presented in this tutorial are designed explicitly to include regularization.

We make a special effort to verify the integrity and robustness of the inversion algorithms presented here.
Our test cases are all noninteracting systems because we can focus on recovering only the known external potential as explained in Sec.~\ref{sec:inversion-examples}.
These systems may seem trivial but their corresponding inverse problems are just as difficult as the inverse problems corresponding to interacting densities and provide great insight into the design of inversion algorithms.
We also strive to use different algorithms and grids for the direct problems in the examples with approximate target densities to avoid any reliance on inverse crimes as described in App.~\ref{app:boundary-conditions}.
The numerical errors in both the inversion methods and the target densities used as input are also tracked carefully to avoid problems with overfitting.
In general, a great deal of numerical analysis is used throughout this work to make sure that our inversion methods are both correct and efficient.


One possible motivation for studying the inverse problem of \gls{dft} is to benchmark the many approximate exchange-correlation functionals currently available.
Benchmarking is usually done on model systems where the density is either known analytically or found numerically through a direct solution of the electronic Schr\"{o}dinger equation.
As seen in many works on inverse problems in \gls{dft},\cite{zhao_electron_1994,van_leeuwen_exchange-correlation_1994,wagner_reference_2012,wu_direct_2003} major features of the exact exchange-correlation potential are often found to be missing in popular approximations and thus the inverse problem can help guide the development of new approximations.

Perhaps an even more practical reason to study the inverse problem of \gls{dft} comes from its connection with the \gls{oep} method.
In the \gls{oep} method, the exchange-correlation functional is an explicit functional of the \gls{ks} orbitals and therefore an implicit functional of the density.\cite{kummel_optimized_2003}
As pointed out in Ref.~\citenum{wu_direct_2003}, the optimization procedures used in the \gls{oep} method can be adapted to perform density-to-potential inversions and vice versa.
This connection with the \gls{oep} method shows that density-to-potential inversions are also important predictive tools in addition to their use in benchmarking or guiding the construction of potentials.

\mysection{\sffamily \Large Direct and Inverse Methods}\label{sec:dft-theory}
We begin our tutorial by highlighting the differences between the forward and inverse problems of \gls{dft}.
Although the direct and inverse problems of \gls{dft} share the same \gls{ks} equations, there are several important differences between them that lead to very different algorithms and convergence criteria.
We briefly describe common numerical methods for solving the direct problem of \gls{dft} and then describe in detail several inverse-problem methods.
All of the examples presented are one-dimensional systems for simplicity but the formulas are written for multidimensional systems to aid future development.
We also limit our studies to spin-compensated systems and drop the spin index on most formulas.

The \gls{ks} equations for a closed-shell system with $2N$ electrons can be written as
\begin{subequations}\label{eq:ks-eqs-direct}
\begin{align}
\varepsilon_{j}\underline{\phi_{j}\!\left(\mathbf{r}\right)} & =\left[-\frac{\Delta}{2}+v_{\text{KS}}\!\left(\left[\underline{n}\right],\mathbf{r}\right)\right]\underline{\phi_{j}\!\left(\mathbf{r}\right)},\label{eq:ks-eval-direct}\\
\underline{n\!\left(\mathbf{r}\right)} & =2\sum_{j=1}^{N}\left|\underline{\phi_{j}\!\left(\mathbf{r}\right)}\right|^{2},\label{eq:ks-density-direct}\\
v_{\text{KS}}\!\left(\left[\underline{n}\right],\mathbf{r}\right) & =v_{\text{ext}}\!\left(\mathbf{r}\right)+v_{\text{H}}\!\left(\left[\underline{n}\right],\mathbf{r}\right)+v_{\text{xc}}\!\left(\left[\underline{n}\right],\mathbf{r}\right),\label{eq:ks-potential-direct}
\end{align}
\end{subequations}
where the underlined quantities are unknowns, $v_{\text{xc}}=\frac{\delta E_{\text{xc}}\left[n\right]}{\delta n}$
is the exchange-correlation potential, 
\begin{equation}
v_{\text{H}}\!\left(\left[n\right],{\bf r}\right)=\int\text{d}{\bf r}'\,n\!\left({\bf r}'\right)\frac{1}{\left|{\bf r}-{\bf r}'\right|}\label{eq:Hartree-potential}
\end{equation}
is the Hartree potential, the external potential $v_{\text{ext}}\!\left({\bf r}\right)$
is usually the electron-nuclei potential, and the orbitals are
constrained to be orthonormal:
\begin{equation}
\braket{\phi_{j}|\phi_{k}}=\delta_{j,k}=\begin{cases}
1 & \text{if }j=k\\
0 & \text{otherwise}
\end{cases}.\label{eq:dft-orthonormal}
\end{equation}
The orbitals are usually chosen such that their corresponding eigenvalues $\left\{ \varepsilon_{j}\right\} $ form the set of lowest possible eigenvalues.
The \gls{ks} equations for the inverse problem of \gls{dft}
\begin{subequations}\label{eq:ks-eqs-inverse}
\begin{align}
\varepsilon_{j}\underline{\phi_{j}\!\left(\mathbf{r}\right)} & =\left[-\frac{\Delta}{2}+v_{\text{KS}}\!\left(\left[n\right],\mathbf{r}\right)\right]\underline{\phi_{j}\!\left(\mathbf{r}\right)},\label{eq:ks-eval-inverse}\\
n\!\left(\mathbf{r}\right) & =2\sum_{j=1}^{N}\left|\underline{\phi_{j}\!\left(\mathbf{r}\right)}\right|^{2},\label{eq:ks-density-inverse}\\
v_{\text{KS}}\!\left(\left[n\right],\mathbf{r}\right) & =v_{\text{ext}}\!\left(\mathbf{r}\right)+v_{\text{H}}\!\left(\left[n\right],\mathbf{r}\right)+\underline{v_{\text{xc}}\!\left(\left[n\right],\mathbf{r}\right)},\label{eq:ks-potential-inverse}
\end{align}
\end{subequations}
only differ from the direct problem in the unknowns as indicated by underlining.

Equation~\eqref{eq:ks-eval-direct} is a nonlinear eigenvalue problem due to the \gls{ks} potential's density dependence whereas Eq.~\eqref{eq:ks-eval-inverse} is a linear eigenvalue problem.
The inverse problem is still nonlinear but the nonlinearity comes from the definition of the density shown in Eq.~\eqref{eq:ks-density-inverse}.
This subtle difference in nonlinearity means that the mixing schemes prevalent in numerical methods for the direct problem are not needed in solving the eigenvalue problem in the inverse problem.
Additionally, the definition of the density in Eq.~\eqref{eq:ks-density-inverse} puts constraints on the values of the \gls{ks} orbitals when solving the inverse problem that aren't present in the direct problem.
These additional constraints are exploited in the inversion method described in \myref{sec:var-constrained}.
Finally, the nonlinearity in both problems makes the choice of initial guess very important as convergence is not guaranteed in many numerical methods when poor initial guesses are used.\cite{le_bris_computational_2005}

The \gls{ks} equations apply to both finite and extended systems provided that the appropriate boundary conditions are applied.
In the direct problem of \gls{dft} the orbitals are usually constrained to satisfy periodic or box-type (Dirichlet) boundary conditions as dictated by the system being studied.\cite{castro_octopus:_2006}
In the inverse problem the boundary conditions may not be known exactly but the density constraint uniquely determines the orbitals' behavior at the boundaries.
If boundary conditions are imposed then it is important that they agree with the density constraint as explained in App.~\ref{app:boundary-conditions}.
We show in this tutorial the alternative of letting the inversion method enforce the orbital boundary conditions to agree with the density constraint.
This method has the advantage that no detailed knowledge of the boundary conditions is needed and the disadvantage that the values of the potential at the boundaries may need to be discarded after the inversion if the density at the boundaries contains some inherent error.
We call this method \gls{dcbc} and discuss it further in \myref{sec:one-orb}.

As mentioned in \myref{sec:introduction}, uniqueness is a key ingredient in solving both direct and inverse problems.
Although the density in \gls{dft} is a unique quantity in both the forward and inverse problems, the \gls{ks} orbitals that it is built from are not unique.
As seen in Eq.~\eqref{eq:ks-density-direct}, the density is formed from the absolute value squared of the orbitals and therefore the orbitals can only be unique up to a phase factor.
Furthermore, any unitary transformation of the \gls{ks} orbitals will also produce the same density.
The \gls{ks} potential in Eq.~\eqref{eq:ks-potential-direct} is also a source of nonuniqueness as it is only unique up to a constant.\cite{chayes_density_1985}
As seen in \myref{sec:dft-inverse}, most density-to-potential inversion methods force the unknown potential to be unique by imposing restrictions of some form.

Numerical convergence in the direct problem of \gls{dft} involves both a convergence in the iterations to self-consistency and a convergence for the eigenvalue problem solved at each iteration.
This convergence is described in great detail in Ref.~\citenum{le_bris_computational_2005} and we only point out here that the potentials are usually given by analytic formulas.
This means that the \gls{ks} equations can be solved self-consistently and the error in the resulting density will depend mainly on the level of discretization employed and the choice of approximate exchange-correlation functional.
Numerical convergence in the inverse problem of \gls{dft} is also limited by the chosen discretization method but it is the quality of the target density that sets a somewhat surprising limit on convergence as explained in this tutorial.

Although most of the test cases presented in this work have analytic formulas for the target densities, most target densities of real interest are not known analytically.
If the numerical error present in a given target density is not accounted for properly in a density-to-potential inversion then overfitting becomes a problem and unphysical features develop in the recovered potential.
Examples of overfitting and convergence limits related to the relative error of the target density are given in App.~\ref{app:numerical-issues}.

\mysubsection{\sffamily \large Direct-Problem Methods}\label{sec:dft-direct}
A variety of numerical methods for solving the \gls{ks} equations exist and can be placed in essentially three different categories: plane waves and grid methods, localized atomic-(like) orbitals, and atomic sphere methods.~\cite{martin_electronic_2004}
We use the finite difference grid method for our calculations similar to the Octopus \gls{tddft} code.\cite{andrade_real-space_2015}
The finite difference method is capable of producing sharp features that would be difficult to find using other common electronic-structure basis sets.
In this section we review several aspects of the finite difference formulation of \gls{dft} that are necessary for solving the direct problem and also relevant to some of the inverse problem methods.
A more thorough review of the finite difference method applied to \gls{dft} can be found in Ref.~\citenum{beck_real-space_2000}.

The orbitals, potentials, and density in the \gls{ks} equations are represented as discrete points on a grid when using the finite difference method.
The potentials are diagonal in the coordinate representation and the Laplacian is a very sparse matrix operator depending on the approximation used.
An example fourth-order Laplacian for a regularly spaced grid with seven points is
\begin{equation}
\frac{1}{h^{2}}\begin{bmatrix}\frac{15}{4} & -\frac{77}{6} & \frac{107}{6} & -13 & \frac{61}{12} & -\frac{5}{6} & 0\\
\frac{5}{6} & -\frac{5}{4} & -\frac{1}{3} & \frac{7}{6} & -\frac{1}{2} & \frac{1}{12} & 0\\
-\frac{1}{12} & \frac{4}{3} & -\frac{5}{2} & \frac{4}{3} & -\frac{1}{12} & 0 & 0\\
0 & -\frac{1}{12} & \frac{4}{3} & -\frac{5}{2} & \frac{4}{3} & -\frac{1}{12} & 0\\
0 & 0 & -\frac{1}{12} & \frac{4}{3} & -\frac{5}{2} & \frac{4}{3} & -\frac{1}{12}\\
0 & \frac{1}{12} & -\frac{1}{2} & \frac{7}{6} & -\frac{1}{3} & -\frac{5}{4} & \frac{5}{6}\\
0 & -\frac{5}{6} & \frac{61}{12} & -13 & \frac{107}{6} & -\frac{77}{6} & \frac{15}{4}
\end{bmatrix},\label{eq:fd-stencil-O4}
\end{equation}
where $h$ is the grid spacing and sided finite differences are used at the boundaries.
Different orders of approximate finite difference operators can be derived using the algorithm described in Ref.~\citenum{fornberg_generation_1988}.
There are many possible boundary conditions and methods to implement them depending on the problem and accuracy desired.\cite{nielsen_quantum_2014,ladouceur_boundaryless_1996,antoine_review_2008}
Custom meshes can also be used that conform to the geometry of a given system as explained in Ref.~\citenum{castro_octopus:_2006}.
One disadvantage of the finite difference method applied to \gls{dft} is the lack of a variational principle for most approximate finite difference operators.
This means that the approximate energy may actually be smaller than the true energy found by solving the \gls{ks} equations exactly.
The approximate energy will approach the true energy as the grid is refined and/or the accuracy of the finite difference operator is improved but it doesn't have to approach it from above unless the finite difference operator is modified as in Ref.~\citenum{maragakis_variational_2001}.

A variety of numerical algorithms are available for computing the potentials in Eq.~\eqref{eq:ks-potential-direct}.
The Hartree potential $v_{\text{H}}$ shown in Eq.~\eqref{eq:Hartree-potential} satisfies the Poisson equation
\begin{equation}
\Delta v\!\left({\bf r}\right)=-4\pi n\!\left({\bf r}\right)\label{eq:poisson}
\end{equation}
and has been studied extensively.
The fast Fourier transform, fast multipole method, and conjugate gradient methods are just a few of the available algorithms for computing $v_{\text{H}}$; a thorough review of these methods can be found in Ref.~\citenum{garcia-risueno_survey_2014}.
As mentioned in \myref{sec:introduction}, there are many approximate exchange-correlation functionals and corresponding potentials.
The \textsc{libxc} library contains a wide selection of these approximations and is easily incorporated in \gls{dft} programs.\cite{marques_libxc:_2012}
Exchange-correlation approximations can also be derived for systems with custom electron-electron interactions as shown in Refs.~\citenum{baker_one-dimensional_2015} and \citenum{wagner_reference_2012}.

\begin{figure}[h]
\begin{center}
\includegraphics[width=0.65\textwidth]{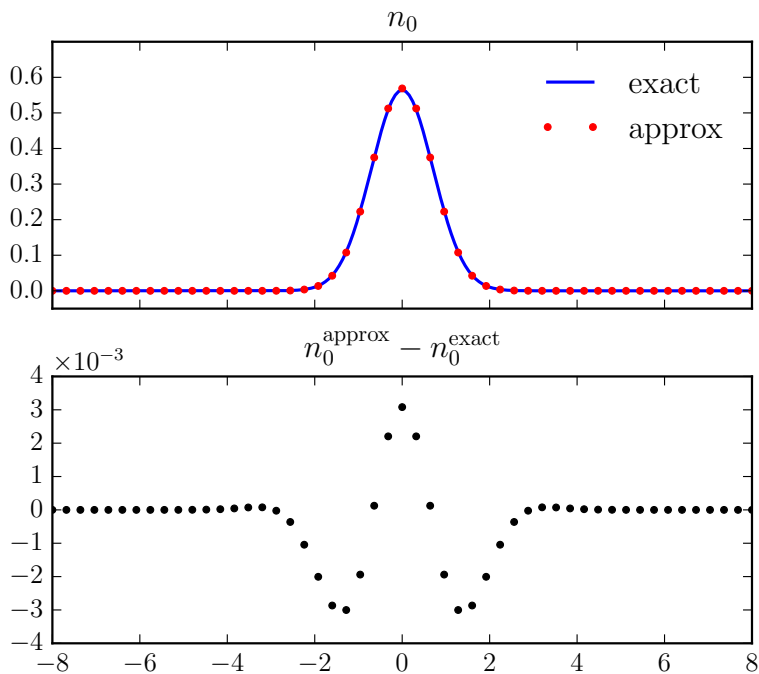}
\end{center}
\caption[The error in the ground state density of the harmonic oscillator when using a finite difference scheme and the \textsc{eig\_banded} routine.]{\label{fig:sho_fd_err} The ground state density of the harmonic oscillator compared to a numerical approximation using the \textsc{eig\_banded} routine and a second-order finite difference approximation to the Laplacian (top).  The majority of the error is in the high-density central region (bottom).}
\end{figure}

\begin{figure}[h]
\begin{center}
\includegraphics[width=0.65\textwidth]{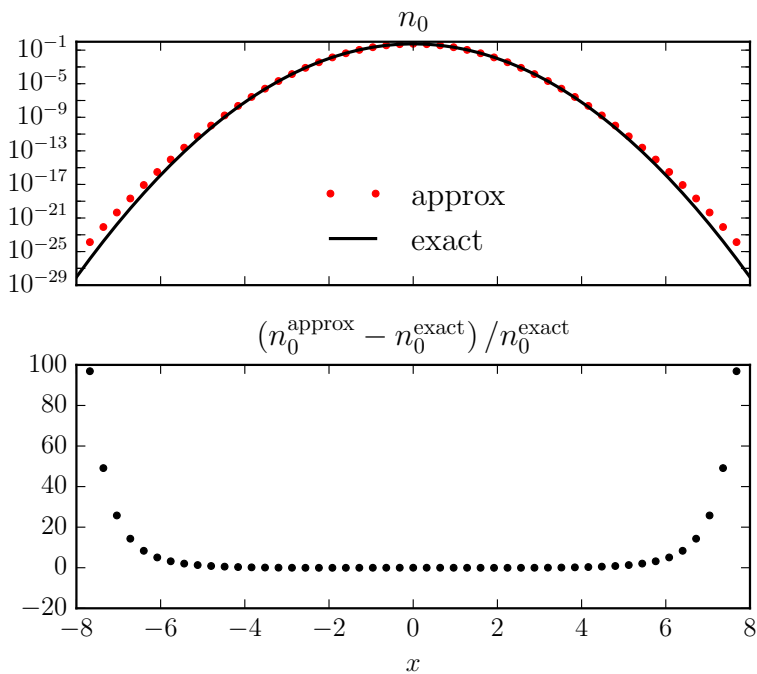}
\end{center}
\caption[The relative error in the ground state density of the harmonic oscillator when using a finite difference scheme and the \textsc{eig\_banded} routine.]{\label{fig:sho_fd_relerr} The same densities displayed in Fig.~\ref{fig:sho_fd_err} are displayed on a logarithmic scale (top).  The relative error in the asymptotic regions is orders of magnitude larger than the relative error in the central region (bottom).}
\end{figure}

The \gls{ks} equations are usually solved using sparse matrix eigenvalue solvers in contrast to the dense solvers typically used in basis-set methods.\cite{beck_real-space_2000}
The examples in this work use the \texttt{eigs} and \texttt{eig\_banded} eigenvalue solvers in the SciPy library, which are interfaces to routines in the \textsc{lapack} and \textsc{arpack} libraries respectively.\cite{scipy-library}
These solvers are capable of computing the eigenvalues and eigenvectors of a sparse matrix to machine precision and are very efficient when only a few eigenvalues and eigenvectors are needed.
If the potentials used in a given calculation are exact and the solvers are allowed to converge to machine precision then the majority of the numerical error will come from the discretization of the Laplacian.
An example of this error is shown in Fig.~\ref{fig:sho_fd_err} for the ground state density of the harmonic oscillator using a second-order approximate Laplacian.
Although the majority of the error is in the central region, Fig.~\ref{fig:sho_fd_relerr} shows that the relative error is actually largest in the asymptotic regions.
The approximate finite difference operators are very accurate in regions where the orbitals are well represented by polynomials and are less accurate in the asymptotic region where the orbitals are dominated by exponential decay.
The large relative error is largely inconsequential for most direct \gls{dft} problems with almost no effect on the total energies but it does play a strong role in the inverse problem as shown in \myref{sec:dft-inverse}.
This example also shows that the errors in most direct \gls{dft} problems are systematic rather than random  noise typical of experimental data and must also be accounted for in the inverse problem as explained in App.~\ref{app:numerical-issues}.

The previous example of a noninteracting system only required solving one eigenvalue problem but a typical \gls{dft} calculation for an interacting system requires multiple eigenvalue solves in what is known as the \gls{scf} method.
The initial density guess for a given system is inserted into the right-hand side of Eq.~\eqref{eq:ks-eval-direct}, the eigenvalue problem is solved for a new set of orbitals, the orbitals are inserted into Eq.~\eqref{eq:ks-density-direct} to produce a new density, and the cycle repeats until a convergence criterion is reached.
This method is usually modified to avoid oscillating densities by mixing the new density with previous densities via the \gls{diis} method\cite{pulay_improved_1982}) or some other form of mixing.\cite{castro_octopus:_2006}
Often the \gls{scf} loop is stopped when the change in density or the change in each eigenvalue becomes smaller than a prescribed value but this does not guarantee global convergence.
An alternative gradient-based method that avoids the repeated eigenvalue computation is also possible as illustrated in Ref.~\citenum{zhang_gradient_2014}.
A more detailed analysis of numerical convergence in \gls{dft} calculations can be found in Ref.~\citenum{le_bris_computational_2005}.

\mysubsection{\sffamily \large Inverse-Problem Methods}\label{sec:dft-inverse}
As is the case with the direct problem, there exist an assortment of numerical methods for solving the inverse  problem of \gls{dft}, each with its own set of advantages and disadvantages.
Here we review several density-to-potential algorithms and introduce two new inversion methods.
All of the methods are presented using the finite-difference method and may include small adaptations made to convert algorithms that were originally designed for basis-set methods.

\mysubsubsection{\sffamily \normalsize One and Two Electrons: One-Orbital Formula}\label{sec:one-orb}
The one-orbital inversion formula
\begin{align}
v_{\text{KS}}\!\left(\left[n\right],\mathbf{r}\right) & =\frac{\Delta\phi_{0}\!\left(\mathbf{r}\right)}{2\phi_{0}\!\left(\mathbf{r}\right)}=\frac{\Delta\sqrt{n_{0}\!\left(\mathbf{r}\right)}}{2\sqrt{n_{0}\!\left(\mathbf{r}\right)}}\label{eq:one-orbital-ti}
\end{align}
can be derived simply by setting the energy to zero and solving for the \gls{ks} potential in Eq.~\eqref{eq:ks-eqs-direct} in terms of the ground state density $n_0$.
(We are allowed to set the energy to zero because the potential is only unique up to a constant.)
This formula is exact for one electron or two electrons with opposite spins in the same orbital and is often referred to as the bosonic or one-electron potential.\cite{ryabinkin_determination_2012}
Although at first glance the one-orbital formula appears somewhat trivial, it is helpful in unraveling many of the numerical problems common to all density-to-potential algorithms and can even serve as a useful approximation for systems with many electrons.

\begin{figure}[h]
\begin{center}
\includegraphics[width=0.65\textwidth]{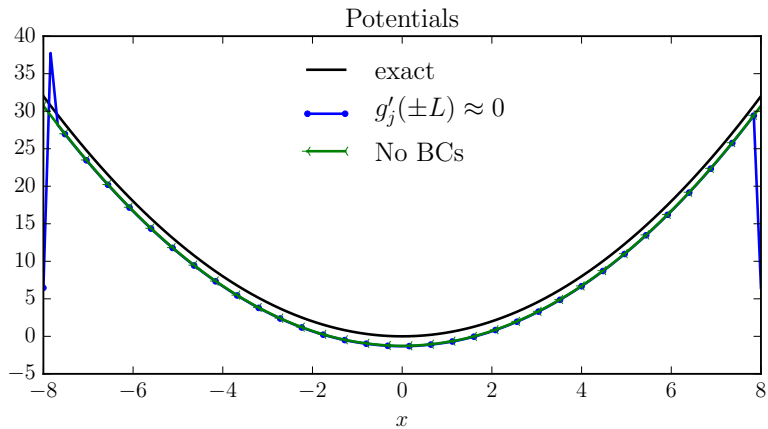}
\end{center}
\caption[The negative effect of enforcing zero derivative boundary conditions on the scaled orbitals in a PDE-constrained optimization.]{\label{fig:compare_bcs} The potentials produced in a density-to-potential inversion using the inversion method described in \myref{sec:pde-constrained} and a target density formed from the sum of the ground state and first two excited states of the harmonic oscillator.  The potential is correct at the boundaries when no boundary conditions are applied to the scaled orbitals (No BCs) and incorrect when the first derivative of the scaled orbitals is set to zero at the boundaries ($g_{j}'\!\left(\pm L\right)\approx0$).}
\end{figure}

The one-orbital formula makes it very clear that the \gls{ks} orbital is uniquely determined up to an arbitrary phase by the target density via the equation $\phi_{0}=\sqrt{n_{0}}$.
This means that the values of the orbital in the entire domain of the problem are known including the boundaries.
In the forward problem of \gls{dft} we usually force the orbitals to satisfy certain boundary conditions but in the inverse problem this is not necessary because the target density uniquely determines their behavior at the boundaries up to an arbitrary unitary transformation of the orbitals.
Both Fig.~\ref{fig:mismatch_boundaries} and the $g_{j}'\!\left(\pm L\right)\approx0$ line in Fig.~\ref{fig:compare_bcs} show examples of molecular system inversions with approximate boundary conditions enforced.
In both examples the resulting potentials have several incorrect values near the boundaries where the potential actually enforces the approximate boundary conditions.
In Fig.~\ref{fig:mismatch_boundaries} the potential should actually be
\begin{align}
v\left(x\right)= & \frac{x^{2}}{2}\left[\Theta\left(x-4\right)-\Theta\left(x+4\right)\right]-x\left[\delta\left(x-4\right)+\delta\left(x+4\right)\right]\nonumber \\
 & +\left[\delta'\left(x-4\right)-\delta'\left(x+4\right)\right]/2,\label{eq:zero-bc-sho}
\end{align}
because the density is artificially forced to be zero at the boundaries and the ``incorrect'' potential values near the boundaries are approximations to this discontinuous behavior, where $\delta\left(x\right)$ is the Dirac delta function and $\Theta\left(x\right)$
is the Heaviside function.
When approximate boundary conditions are not enforced in these examples then the potential is correct everywhere including at and near the boundaries.

Although the one-orbital formula is exact for one-electron and most closed-shell two-electron systems, it must be carefully implemented numerically to avoid spurious results.
As seen in App.~\ref{app:target-density-noisy}, noisy densities can easily produce wildly inaccurate potentials when using Eq.~\eqref{eq:one-orbital-ti} and finite difference operators.
Such noisy behavior is unphysical because densities arising from the Schr\"{o}dinger equation are theoretically continuous even when generated by singular potentials.\cite{merzbacher_quantum_1998}
The following examples show how to appropriately smooth target densities based on the noise level before inserting them into the one-orbital formula.

\begin{figure}[h]
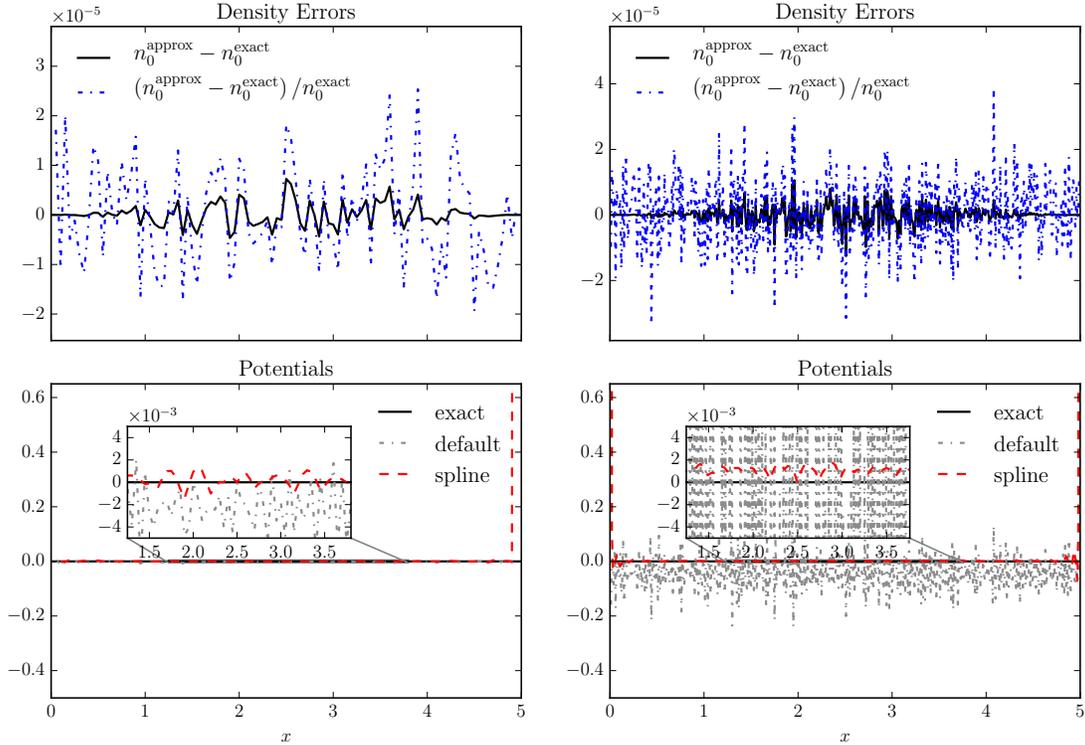

\begin{center}
\includegraphics[width=0.45\textwidth]{\myfigs{piab_smooth101}}\includegraphics[width=0.45\textwidth]{\myfigs{piab_smooth501}}
\end{center}
\caption[A comparison of the one-orbital formula with and without smoothing a noisy target density.]{\label{fig:piab_smooth} The ground-state density of a particle-in-a-box is contaminated with weighted randomly-distributed noise (top) and then inserted into the one-orbital formula before smoothing (default) and after smoothing (spline) via a weighted cubic spline (bottom).  The smoothing has increasing importance as the number of grid points is increased from 101 (left) to 501 (right) even though the target density's error level is the same.}
\end{figure}

We first illustrate smoothing using the particle-in-a-box\cite{merzbacher_quantum_1998} ground state density
\begin{equation}
n_{0}^{\text{exact}}\!\left(x\right)=\frac{2}{L}\sin^{2}\left(\frac{\pi x}{L}\right)\label{eq:particle-box-gstate}
\end{equation}
corresponding to the potential
\begin{equation}
v\!\left(x\right)=\begin{cases}
0 & \text{if }0<x<L\\
\infty & \text{otherwise}
\end{cases}.\label{eq:particle-box-pot}
\end{equation}
If we take the exact ground-state density $n_{0}^{\text{exact}}$ and add a small amount of weighted noise generated from the standard normal distribution 
$\ensuremath{\operatorname{N}\!\left(\mu,\sigma^{2}\right)}$\cite{bain_introduction_2000} according to the formula
\begin{equation}
n_{0}^{\text{noise}}=n_{0}^{\text{exact}}\left[1+\ensuremath{\operatorname{N}\!\left(0,1\times10^{-10}\right)}\right],\label{eq:weighted-noise}
\end{equation}
then the potential resulting from the one-orbital formula will also be noisy with a level proportional to the number of grid points.
This noise can be removed by fitting a cubic spline to the target density weighted by the inverse of the approximate standard deviation $w=1/(\sigma*n_{0}^{\text{approx}})$.
We implement this smoothing procedure using the \texttt{UnivariateSpline} routine in the SciPy library and the results are shown in Fig.~\ref{fig:piab_smooth} for a box of length 5 with varying numbers of grid points.
This example shows that smoothing is essential for dense grids with errors but not as important on coarse grids since the derivative of the error doesn't dominate in the one-orbital formula as explained in App.~\ref{app:finite-diff-operators}.

\begin{figure}[h]
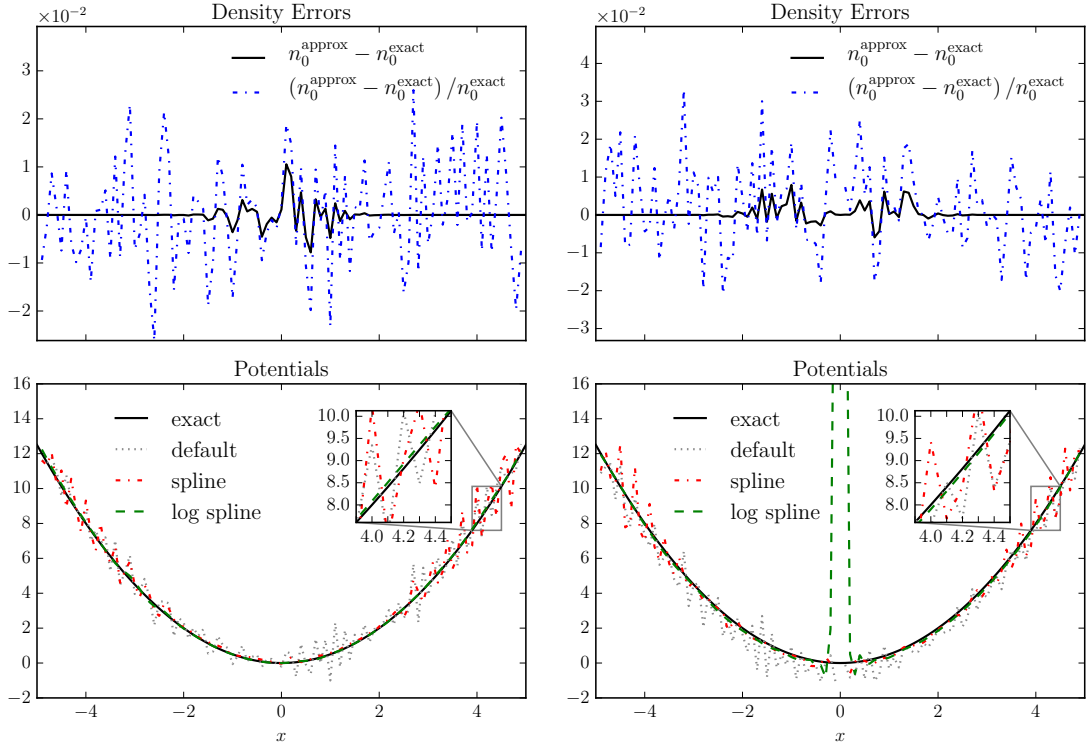

\begin{center}
\includegraphics[width=0.45\textwidth]{\myfigs{sho_smooth_n_0}}\includegraphics[width=0.45\textwidth]{\myfigs{sho_smooth_n_1}}
\end{center}
\caption[A comparison of the one-orbital formula with and without logarithmic scaling before smoothing a noisy target density.]{\label{fig:sho_smooth} The ground-state density of the harmonic oscillator is contaminated with weighted randomly-distributed noise (top left) and then inserted into the one-orbital formula without smoothing (default), with smoothing (spline), and with a logarithmic transform before smoothing (log spline) via a weighted cubic spline (bottom left).  The process is repeated on the right for the density of the first excited state and the distinct delta-well singularity in the potential at $x=0$ is recovered.}
\end{figure}

The target density in the previous example was well approximated by a cubic spline but other target densities may need additional modifications before smoothing is applied.
In particular, exponentially decaying densities are not adequately described by cubic splines unless they are first logarithmically transformed.
We show how this can be done using the ground-state of the harmonic oscillator with the same weighted noise given by Eq.~\eqref{eq:weighted-noise}.
In this case we take the logarithm of the noisy density, fit this new quantity to a cubic spline using the weights $w=1/\log(1+\sigma)$, and then exponentiate the result to get the smoothed target density.
The potential produced using this logarithmic scaling before fitting is much more accurate in the asympototic region than the potential produced with no scaling before smoothing as seen in Fig.~\ref{fig:sho_smooth}.
We refer to this smoothing procedure later in this tutorial as the logarithmic smoothing method.
Even highly singular potentials can be recovered using this method provided that the locations of zero density are first averaged with their nearest neighbors before applying the scaling and smoothing.
(This or a similar procedure must be followed to avoid division-by-zero errors.)
Figure~\ref{fig:sho_smooth} shows the case described in Ref.~\citenum{maitra_demonstration_2001} of treating the density of the first excited state of the simple harmonic oscillator as a ground state density.
The singular feature of the potential ($\delta\!\left(x\right)/\left|x\right|$) becomes more pronounced with this procedure as the grid is refined.

\begin{figure}[h]
\begin{center}
\includegraphics[height=0.7\textheight]{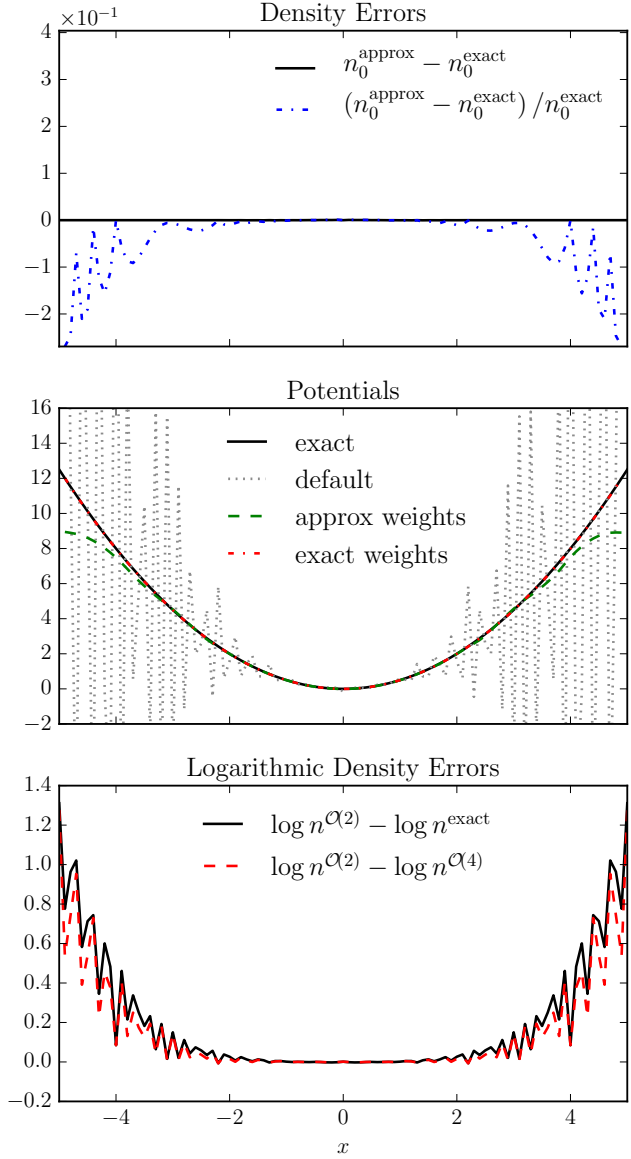}
\end{center}
\caption[A comparison of the one-orbital formula using smoothing with approximate and exact weights.]{\label{fig:sho_err_est} The ground-state density of the harmonic oscillator is computed using an infinite-to-finite spatial grid mapping (top) and then inserted into the one-orbital formula with both approximate and exact scaling weights (middle).  The approximate weights are computed using an error estimate that tends to underestimate the error in the asymptotic region (bottom) and leads to an underestimated potential in that region.}
\end{figure}

In the previous examples we were able to correctly smooth the target densities by having estimates of the standard deviation at each grid point.
Such information would be typical of an experimentally determined density but most of the error patterns encountered in our density-to-potential inversions come from  computationally determined densities using interacting wave function methods.
In these cases we can use error estimates to compute the correct weights for the smoothing procedure.
We show how this can be done by computing the ground state density of the harmonic oscillator using the infinite-to-finite spatial grid mapping of Ref.~\citenum{ladouceur_boundaryless_1996} with a scaling parameter of $\alpha=2$.
(We purposely use a different grid to compute the target density in order to avoid comitting an inverse crime.)
We compute the target density on this grid once with a second-order discretization of the Laplacian and 101 points, (labeled $n^{\mathcal{O}\!\left(2\right)}$), and then again with a fourth-order approximate Laplacian and 201 points, (labeled $n^{\mathcal{O}\!\left(4\right)}$).
These two densities are linearly interpolated to the same equally-spaced grid shown in Fig.~\ref{fig:sho_smooth} and subtracted from each other to form an error estimate at each grid point.
We then apply the logarithmically scaled spline fitting procedure described above with the approximate weights $w=1/\left|\log n^{\mathcal{O}\!\left(2\right)}-\log n^{\mathcal{O}\!\left(4\right)}\right|$
and the exact weights $w=1/\left|\log n^{\mathcal{O}\!\left(2\right)}-\log n^{\mathrm{exact}}\right|$.
The results of this calculation are displayed in Fig.~\ref{fig:sho_err_est} and show that the error pattern of the approximate target density is far more systematic than the random noise error patterns in the previous examples.
The smoothing using approximate weights underestimates the potential at the edges of the box because the error estimate is too small in that region.
If the error estimate is computed using 501 points instead of 201 points in the fourth-order density computation then the potential is recovered correctly in the entire box.
This example illustrates the need not only for error estimates but also for the need to put bounds on the error estimates either through computing additional error estimates or through statistical means.

\begin{figure}[h]
\begin{center}
\includegraphics[width=0.65\textwidth]{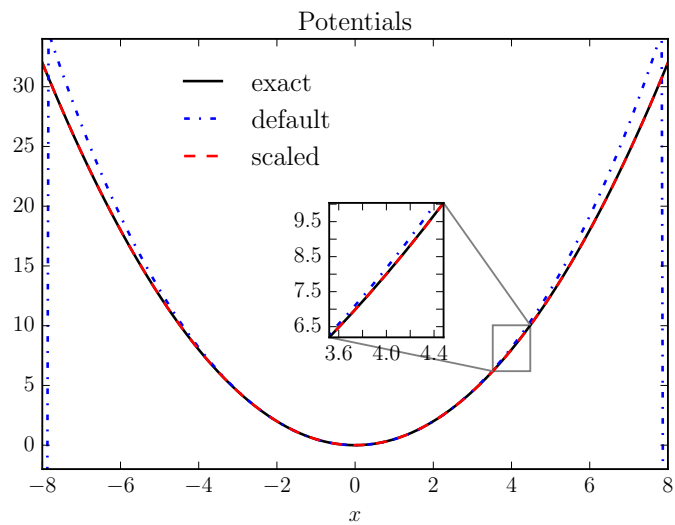}
\end{center}
\caption[A comparison of the one-orbital formula with and without density scaling.]{\label{fig:sho_scaling} The ground-state density of the harmonic oscillator is inserted into the one-orbital formula without scaling (default) and with density scaling (scaled).  The scaled finite difference operator performs much better in the asymptotic region $\left|x\right|>4$ and doesn't have problems near the edges of the box.}
\end{figure}

Even if the target density is correctly smoothed before applying the one-orbital formula, the finite difference operator used to approximate the Laplacian can severely limit the accuracy of a density-to-potential inversion.
As was the case with the spline fitting procedure described above for exponentially decaying densities, the finite difference method performs poorly in the asymptotic region unless extra precautions are taken.
In \myref{sec:pde-constrained} we show how the finite difference operators can be modified to treat much of the exponential decay analytically through a scaling procedure.
Figure~\ref{fig:sho_scaling} shows a comparison of the one-orbital formula applied to the exact ground state of the harmonic oscillator with and without scaling the finite difference operator.

\begin{figure}[h]
\begin{center}
\includegraphics[width=0.65\textwidth]{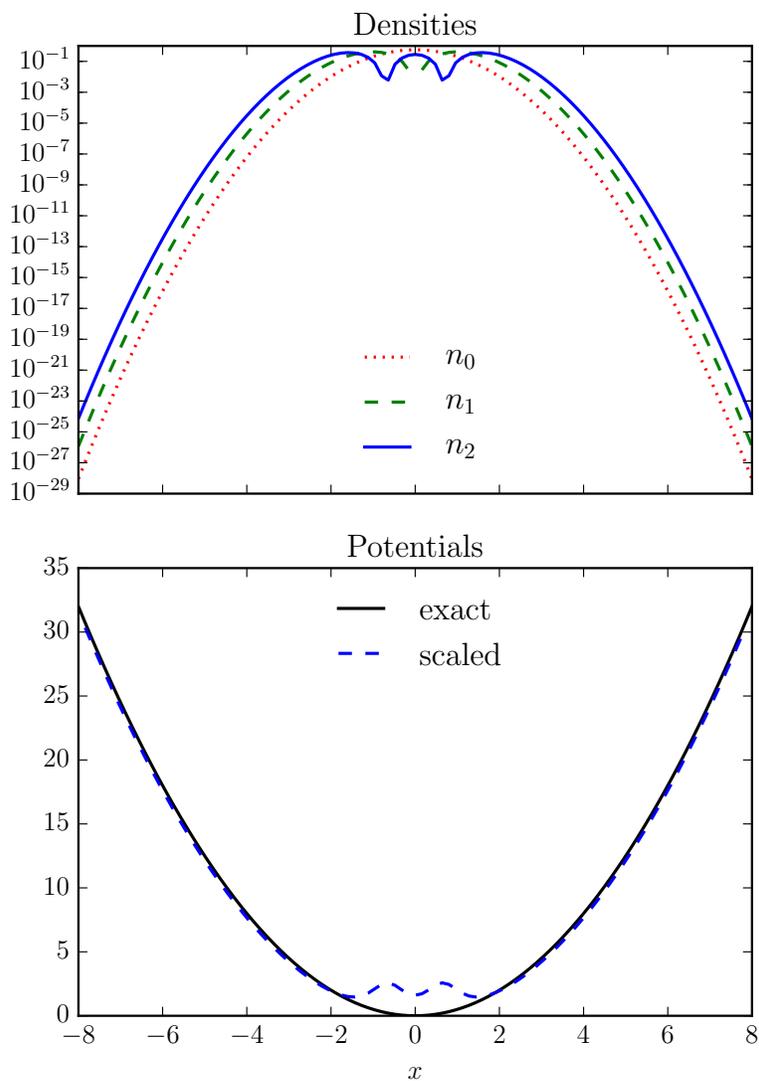}
\end{center}
\caption[The one-orbital formula used as an approximation for a noninteracting system consisting of multiple orbitals.]{\label{fig:sho_one_orb_approx} The densities of the ground-state and first two excited states of the harmonic oscillator (top) are summed and inserted into the one-orbital formula with density scaling (scaled).  The resulting potential is an excellent approximation in the region $\left|x\right|>2$ where the second excited state density is the dominant contribution to the total density.}
\end{figure}

Although the one-orbital formula is only exact for one- or two-electron systems, it is a remarkably good approximation for systems with large regions where the target density is dominated by one orbital.
We illustrate the power of this approximation by applying it to a target density formed from the sum of the ground state and first two excited states of the harmonic oscillator.
The harmonic potential is correctly recovered in the asymptotic region where most of the density comes from the second excited state as seen in Fig.~\ref{fig:sho_one_orb_approx}.
The potential in the central region is not correct but it is a smooth approximation and is the same order of magnitude as the correct potential.
Furthermore, the cost of this approximation is simply one sparse matrix-vector multiplication followed by an array division, which is negligible in comparison to the many sparse matrix-vector multiplications typically needed to solve the eigenvalue problem in Eq.~\eqref{eq:ks-eval-inverse}.

The examples in this section demonstrate how the convergence criteria for \gls{dft} inverse problems are different from those of the direct problem.
These examples and App.~\ref{app:numerical-issues} clearly show that the size and location of the relative error in the target density is the main constraint on the accuracy of the potential resulting from a density-to-potential inversion.
In many cases the target density must first be smoothed before performing an inversion and its standard deviation is the key ingredient in performing this smoothing operation correctly.
We chose to use a spline-based smoothing filter because it is easy to incorporate the standard deviation but other methods can be used with similar success such as the Savitzky-Golay filter.\cite{savitzky_smoothing_1964_3}
The accuracy of the finite difference operators also puts a strict limit on the quality of the recovered potential as some operators simply aren't good approximations in certain regions of rapidly changing density.
Furthermore, the computational cost of the inverse problem using the one-orbital formula is much lower in these examples than in the forward problem and can serve as a useful approximation for many-electron systems.

\mysubsubsection{\sffamily \normalsize Standard Methods}\label{sec:previous-methods}
A number of density-to-potential inversion methods have been developed besides the one-orbital inversion formula.
In this section we look at the iterative inversion method of Ref.~\citenum{van_leeuwen_exchange-correlation_1994} and the direct optimization method of Ref.~\citenum{wu_direct_2003}.
A more complete review of the available density-to-potential inversion methods is found in Ref.~\citenum{wu_direct_2003}.
The two inversion schemes described in this section are fairly representative of the available methods and serve as useful benchmarks later in this chapter.
The target density in the remainder of this work is written as $\tilde{n}$ in order to distiguish it from the intermediate densities created in the iterative procedures that we describe here.
When it does not lead to confusion, we also drop the subscript KS on the \gls{ks} potential $v_{\text{KS}}\!\left(\mathbf{r}\right)$ and the subscript orbs on the number of orbitals $N_{\text{orbs}}$ to simplify the presentation.

{\sffamily \small Van Leeuwen and Baerends:}
The iterative inversion method of van Leeuwen and Baerends described in Ref.~\citenum{van_leeuwen_exchange-correlation_1994} is derived by multiplying both sides of Eq.~\eqref{eq:ks-eval-inverse} by $\phi_{j}^{*}$, summing over $j$, and then dividing by the density given in Eq.~\eqref{eq:ks-density-inverse} to produce the formula
\begin{equation}
v\!\left(\mathbf{r}\right)=\frac{2}{n\!\left(\mathbf{r}\right)}\sum_{j=1}^{N}\phi_{j}^{*}\!\left(\mathbf{r}\right)\frac{\Delta}{2}\phi_{j}\!\left(\mathbf{r}\right)+\varepsilon_{j}\left|\phi_{j}\!\left(\mathbf{r}\right)\right|^{2}.\label{eq:leeuwen-vks}
\end{equation}
This formula is then turned into an iterative scheme by placing the target density in the denominator and iterating until convergence via the formula
\begin{align}
v^{k+1}\!\left(\mathbf{r}\right) & =\frac{2}{\tilde{n}\!\left(\mathbf{r}\right)}\sum_{j}^{N}\frac{1}{2}\phi_{j}^{k*}\!\left({\bf r}\right)\Delta\phi_{j}^{k}\!\left(\boldsymbol{r}\right)+\epsilon_{j}^{k}\left|\phi_{j}^{k}\!\left({\bf r}\right)\right|^{2}\nonumber \\
 & =\frac{1}{\tilde{n}\!\left({\bf r}\right)}\left[n^{k}\!\left({\bf r}\right)v^{k}\!\left({\bf r}\right)\right]=\frac{n^{k}\!\left({\bf r}\right)}{\tilde{n}\!\left({\bf r}\right)}v^{k}\!\left({\bf r}\right),\label{eq:leeuwen-viter}
\end{align}
where $k$ is the iteration index.
This iterative scheme simply increases $v^{k+1}\!\left(\mathbf{r}\right)$ in regions where $n^{k}\!\left({\bf r}\right)>\tilde{n}\!\left({\bf r}\right)$ and decreases it in regions where $n^{k}\!\left({\bf r}\right)<\tilde{n}\!\left({\bf r}\right)$ as seen in the last line of Eq.~\eqref{eq:leeuwen-viter}.
The iterations are stopped when 
\begin{equation}
\max_{{\bf r}}\left|1-\frac{n^{k}\!\left({\bf r}\right)}{\tilde{n}\!\left({\bf r}\right)}\right|<\epsilon\label{eq:leeuwen-converg}
\end{equation}
for some desired threshold $\epsilon$.
This scheme is sensitive to the initial potential guess $v^{0}\!\left(\mathbf{r}\right)$ and usually requires a prefactor to avoid wild potential oscillations.
Reference~\citenum{van_leeuwen_exchange-correlation_1994} suggests choosing this prefactor by enforcing the condition
\begin{equation}
1-\delta<\gamma\frac{n^{0}\!\left({\bf r}\right)}{\tilde{n}\!\left({\bf r}\right)}<1+\delta,\label{eq:leeuwen-prefactor}
\end{equation}
where $\delta\approx0.05$ and $\gamma$ is the prefactor.

This iterative inversion scheme of van Leeuwen and Baerends is very easy to implement as it only requires a \gls{pde} solver to produce the new density $n^{k}\!\left({\bf r}\right)$ at each iteration.
We do, however, make some modifications to the original scheme by not forcing the potential to approach any given limit at the boundaries and by using the formula
\begin{equation}
v^{k+1}\!\left(\mathbf{r}\right)=v^{k}\!\left({\bf r}\right)+\gamma\frac{n^{k}\!\left({\bf r}\right)-\tilde{n}\!\left({\bf r}\right)}{\tilde{n}\!\left({\bf r}\right)}\label{eq:leeuwen-modified}
\end{equation}
instead of Eq.~\eqref{eq:leeuwen-viter}.
We don't enforce boundary conditions on the potential because many of the test cases we use recover the external potential with unknown boundary conditions as opposed to just the exchange-correlation potential with its known asymptotic behavior.\cite{van_leeuwen_exchange-correlation_1994}
(Note that the orbitals may still have boundary conditions and the above discussion applies only to fixing values of the potential at the boundaries.)
We use Eq.~\eqref{eq:leeuwen-modified} instead of Eq.~\eqref{eq:leeuwen-viter} because it still uses the same principle of increasing (decreasing) the potential in regions where the density is too large (small) but does not need boundary conditions to function properly.
Other modifications are also possible such as the use of an approximate density-density response matrix to help guide the iterations.\cite{wagner_kohn-sham_2014}

This method produces smooth potentials by construction provided that the underlying eigenvalue solver produces smooth orbitals.
The reliance on a prefactor, however, requires some manual intervention or heuristics to ensure convergence.
Large prefactors may result in wildly oscillating potentials that never converge while small prefactors may produce very small potential changes that take an inordinate amount of time to converge.
Choosing a prefactor is usually not a problem for isolated inversions but can become very tedious when repeatedly applying the procedure to cases with many similar inversions such as in the creation of dissociation curves.\cite{nafziger_fragment-based_2015}


{\sffamily \small Wu-Yang Algorithm}
The Wu-Yang inversion algorithm first introduced in Ref.~\citenum{wu_direct_2003} is one of the more sophisticated and versatile \gls{dft} inversion methods.
In this inversion algorithm the Levy constrained-search formulation of \gls{dft} is modified to produce an unconstrained optimization method.
The functional
\begin{equation}
W_{s}\!\left[v\!\left({\bf r}\right)\right]=2\sum_{j=1}^{N}\Braket{\phi_{j}|\hat{T}|\phi_{j}}+\int\text{d}{\bf r}\,v\!\left({\bf r}\right)\left[n\!\left({\bf r}\right)-\tilde{n}\!\left({\bf r}\right)\right]\label{eq:wy-functional}
\end{equation}
is maximized to produce the optimal potential corresponding to the target density $\tilde{n}\!\left({\bf r}\right)$.
(The kinetic energy is doubled because we are still assuming the orbitals are doubly occupied.)
This maximization is usually performed using classical optimization routines and is very efficient because the required first derivatives (Jacobian values) are given by the simple formula
\begin{equation}
\frac{\delta W_{s}\!\left[v\!\left({\bf r}\right)\right]}{\delta v\!\left({\bf r}\right)}=n\!\left({\bf r}\right)-\tilde{n}\!\left({\bf r}\right).\label{eq:wy-deriv1}
\end{equation}
There is very little extra computational effort required to compute these derivatives because the density is already computed at each iteration while computing Eq.~\eqref{eq:wy-functional}.
The second derivatives (Hessian values) can also be computed either through first-order perturbation theory\cite{wu_direct_2003} or via the discrete adjoint-method described in App.~\ref{app:discrete-adjoint}.

The Wu-Yang inversion algorithm was developed originally for the \gls{oep} method of \gls{dft} and later modified for the \gls{dft} inverse problem.\cite{yang_direct_2002,wu_direct_2003}
The method is efficient and robust when implemented properly but can suffer from rounding errors as shown at the end of App.~\ref{app:numerical-issues}.
It can also produce unphysical highly oscillatory potentials when an unbalanced basis set is used without regularization.\cite{bulat_optimized_2007}

\mysubsubsection{\sffamily \normalsize PDE-Constrained Optimization}\label{sec:pde-constrained}
The inversion schemes described in the previous section rely on modifications to the \gls{ks} equations or underlying energy functionals.
Such modifications are not essential in creating \gls{dft} inversion methods as we show in this section through the use of \gls{pde}-constrained optimization.
Although conceptually simpler than many other inversion procedures, \gls{pde}-constrained optimization can be difficult to implement due to the large number of unknowns and programming challenges involved in the optimization procedure.\cite{jensen_numerical_2016}
In this section we give a brief introduction to \gls{pde}-constrained optimization in the context of \gls{dft} inverse problems and address many of these challenges.
We also introduce an important scaling concept that plays a critical role in increasing the accuracy of the potential in the asymptotic regions of \gls{dft} inverse problems.

\Gls{pde}-constrained optimization applied to the \gls{dft} inverse problem amounts to a fitting procedure in which the unknown potential is optimized until the corresponding density matches the target density.
We use the weighted least-squares cost functional
\begin{equation}
F\!\left[v\right]=\frac{1}{2}\left\Vert \sqrt{w\!\left({\bf r}\right)}\left[n\!\left(\left[v\right],{\bf r}\right)-\tilde{n}\!\left({\bf r}\right)\right]\right\Vert _{2}^{\,2}\label{eq:cost-leastsqrs}
\end{equation}
as our measure of fitness, where the subscript 2 indicates the $L^2$ norm and $w\!\left({\bf r}\right)$ is a positive definite weighting function.\cite{vogel_analysis_1995}
The density $n\!\left(\left[v\right],{\bf r}\right)$ comes from solving Eqs.~\eqref{eq:ks-eval-inverse} and \eqref{eq:ks-density-inverse} for a given potential chosen during the optimization procedure.
The numerical minimization of the discretized Eq.~\eqref{eq:cost-leastsqrs} usually involves a very large number of unknown potential values and generally requires the use of gradient/Jacobian-based optimization algorithms.\cite{biegler_large-scale_2003}
The derivation, programming, and computation of these gradients, [functional derivatives of the cost functional given by Eq.~\eqref{eq:cost-leastsqrs}], is the main difficulty in applying \gls{pde}-constrained optimization to the \gls{dft} inverse problem.

We use the discrete adjoint-method to compute the cost functional derivatives of Eq.~\eqref{eq:cost-leastsqrs}.
This method is also called the discretize-then-differentiate method because the cost functional is first discretized and then differentiated with respect to the potential.\cite{hicken_dual_2014}
Appendix~\ref{app:discrete-adjoint} contains a complete derivation of the discrete adjoint equations for the \gls{dft} inverse problem in multiple dimensions.
These derivatives are then employed in the \gls{tn} algorithm of Ref.~\citenum{nash_newton-type_1984} as implemented in the SciPy library\cite{scipy-library} to optimize the unknown potential until a desired density fit is found.
The cost of computing the functional derivatives via the discrete adjoint-method is roughly the same as computing $F\!\left[v\right]$ itself depending on the choice of linear solver.

The \gls{pde}-constrained inversion method outlined above can also suffer from rounding errors similar to the Wu-Yang algorithm described in the previous section.
These errors can be avoided simply by weighting the densities in the asymptotic region more heavily than in the high-density regions.
One possibility is to set $w\!\left({\bf r}\right)=1/\left[\tilde{n}\!\left({\bf r}\right)\right]^{2}$ so that the relative error is minimized instead of the absolute error.
This weighting scheme in combination with the rescaling described below allows us to accurately reproduce the unknown potential in all regions of a given density-to-potential problem.

\begin{figure}[h]
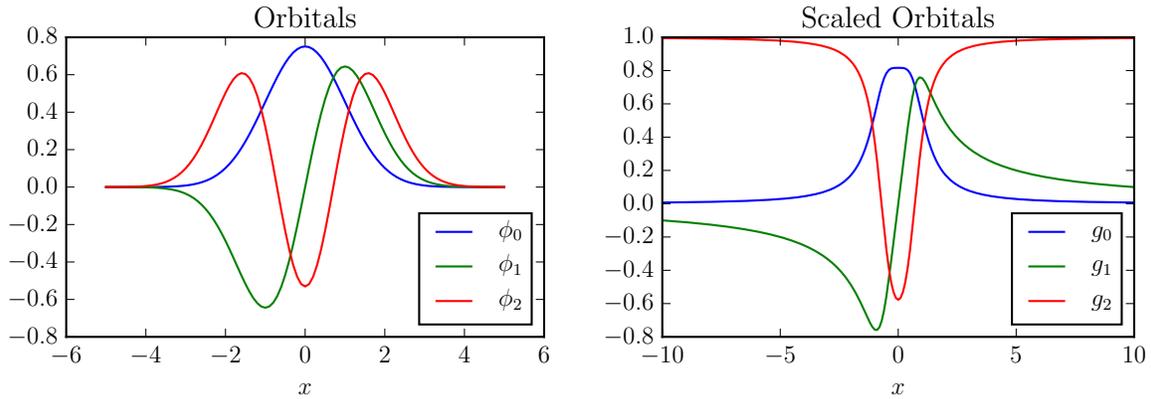

\begin{center}
\includegraphics[width=0.48\textwidth]{\myfigs{sho_orbs}}\includegraphics[width=0.48\textwidth]{\myfigs{scaled_sho_orbs}}
\end{center}
\caption[The first three orbitals of the one-dimensional simple harmonic oscillator with and without density scaling.]{\label{fig:sho_orbs_scaled} The first three orbitals of the one-dimensional simple harmonic oscillator without scaling (left) and with scaling by the square root of the molecular density (right).}
\end{figure}

As mentioned in \myref{sec:one-orb}, the finite difference operators used to approximate the Laplacian are not very accurate in regions where the exponential decay dominates the behavior of the \gls{ks} orbitals.
The finite difference operators would be accurate in all regions if we could apply them to the logarithm of the orbitals similar to the logarithmic spline fitting procedure described earlier.
The \gls{ks} orbitals, however, are not positive semidefinite like the particle density so we instead write them as $\phi_{j}\!\left(\mathbf{r}\right)=\sqrt{\tilde{n}\!\left({\bf r}\right)}g_{j}\!\left(\mathbf{r}\right)$ and solve for the scaled orbitals $\left\{ g_{j}\!\left(\mathbf{r}\right)\right\} $ in the \gls{ks} equations
\begin{subequations}\label{eq:ks-eqs-rescaled}
\begin{align}
\varepsilon_{j}\sqrt{\tilde{n}\!\left({\bf r}\right)}g_{j}\!\left(\mathbf{r}\right) & =\left[-\frac{\Delta}{2}+v\!\left(\mathbf{r}\right)\right]\sqrt{\tilde{n}\!\left({\bf r}\right)}g_{j}\!\left(\mathbf{r}\right)\text{ and}\label{eq:ks-eval-rescaled}\\
n\!\left(\mathbf{r}\right) & =\tilde{n}\!\left({\bf r}\right)\sum_{j=1}^{N}\left|g_{j}\!\left(\mathbf{r}\right)\right|^{2}.\label{eq:ks-density-rescaled}
\end{align}
\end{subequations}
After distributing the Laplacian and canceling common factors, Eq.~\eqref{eq:ks-eval-rescaled} can be rewritten as
\begin{equation}
\varepsilon_{j}g_{j}\!\left(\mathbf{r}\right)=-\frac{1}{2}\left[\frac{\Delta\sqrt{\tilde{n}\!\left({\bf r}\right)}}{\sqrt{\tilde{n}\!\left({\bf r}\right)}}+2\frac{\boldsymbol{\nabla}\sqrt{\tilde{n}\!\left({\bf r}\right)}}{\sqrt{\tilde{n}\!\left({\bf r}\right)}}\cdot\boldsymbol{\nabla}+\Delta+v\!\left(\mathbf{r}\right)\right]g_{j}\!\left(\mathbf{r}\right).\label{eq:ks-eval-expanded}
\end{equation}
The scaling factor can also be written as
\begin{equation}
\sqrt{\tilde{n}\!\left({\bf r}\right)}=\exp\left\{ \frac{1}{2}\log\tilde{n}\!\left({\bf r}\right)\right\} \label{eq:scale-fac}
\end{equation}
and differentiated to produce the terms
\begin{align}
\frac{\boldsymbol{\nabla}\sqrt{\tilde{n}\!\left({\bf r}\right)}}{\sqrt{\tilde{n}\!\left({\bf r}\right)}} & =\frac{1}{2}\boldsymbol{\nabla}\log\tilde{n}\!\left({\bf r}\right)\text{ and}\label{eq:scaled-gradient}\\
\frac{\Delta\sqrt{\tilde{n}\!\left({\bf r}\right)}}{\sqrt{\tilde{n}\!\left({\bf r}\right)}} & =\frac{1}{2}\left[\frac{1}{2}\left|\boldsymbol{\nabla}\log\tilde{n}\!\left({\bf r}\right)\right|^{2}+\Delta\log\tilde{n}\!\left({\bf r}\right)\right].\label{eq:scaled-laplacian}
\end{align}
The derivatives of $\log\tilde{n}\!\left({\bf r}\right)$ and $\left\{ g_{j}\!\left(\mathbf{r}\right)\right\} $ are well approximated by the finite difference operators described in \myref{sec:dft-direct} because they do not decay exponentially.
An example of the scaled orbitals for a system consisting of the first three orbitals of the harmonic oscillator can be seen in Fig.~\ref{fig:sho_orbs_scaled}.

Although it is common to enforce boundary conditions on the orbitals when solving the \gls{ks} equations, we simply let the numerical optimization routine choose the ideal boundary conditions via the \gls{dcbc} method described in \myref{sec:one-orb}.
In Fig.~\ref{fig:sho_orbs_scaled}, the scaled orbitals appear to have derivatives of nearly zero near the boundaries so it may seem reasonable to apply zero derivative boundary conditions when using scaled orbitals.
This assumption leads to errors near the boundaries as shown in Fig.~\ref{fig:compare_bcs} in which the \gls{pde}-constrained optimization procedure outlined above is applied to a target density formed from the sum of the ground state and first two excited states of the harmonic oscillator.
This example clearly shows that enforcing approximate zero derivative boundary conditions on the scaled orbitals results in the incorrect potential at the boundary whereas the potential is recovered correctly when no boundary conditions are applied.
If the target density happens to be produced using artificial boundary conditions then the potential recovered when not applying boundary conditions on the orbitals may have several incorrect values at the edge of the simulation box.

\mysubsubsection{\sffamily \normalsize Constrained Variational Method}\label{sec:var-constrained}
The previous inversion methods described in this chapter all require the solution of an eigenvalue problem at each iteration.
In the case of small systems this is not a severe constraint but as the system size increases the solution of the eigenvalue problem begins to dominate the computation.\cite{le_bris_computational_2005}
Although dismissed in Ref.~\citenum{wu_direct_2003} as ``difficult to perform'', the Levy constrained-search formulation can be used directly to perform a density-to-potential inversion without repeatedly solving eigenvalue problems at each iteration.
In this section we show how to overcome many of the difficulties involved in this constrained-search inversion method through the use of scaling and regularization.

In the Levy constrained-search formulation of \gls{dft}, the noninteracting kinetic energy is minimized subject to the constraints that the orbitals are orthonormal and produce the molecular density of the real system.
The functional to be minimized in this search can be written using Lagrange multipliers as
\begin{multline}
J\left(\left\{ \phi_{j}\right\} \right)=\int\text{d}{\bf r}\left\{ \sum_{j=1}^{N_{\text{orbs}}}\left|{\bf \boldsymbol{\nabla}}\phi_{j}\right|^{2}+v_{\text{KS}}\!\left({\bf r}\right)\left[\sum_{j=1}^{N_{\text{orbs}}}\phi_{j}^{2}\!\left({\bf r}\right)-\tilde{n}\!\left({\bf r}\right)\right]\right.\\
\left.+\sum_{j=1}^{N_{\text{orbs}}}\sum_{k=j}^{N_{\text{orbs}}}\varepsilon_{j,k}\left[\phi_{j}\!\left({\bf r}\right)\phi_{k}\!\left({\bf r}\right)-\delta_{j,k}\right]\right\} ,\label{eq:levy-constrained}
\end{multline}
where the orbitals are assumed to be real for simplicity of presentation.
The first term in Eq.~\eqref{eq:levy-constrained} is one possible form of the kinetic energy and lacks a factor of $1/2$ because the orbitals are doubly occupied.
This expression can be used directly in a constrained numerical optimization program to find the orbitals and \gls{ks} potential for a given target density without the need to solve an eigenvalue problem at each iteration.
The orbitals themselves are being varied by the optimizer to minimize the kinetic energy while satisfying the density and orthonormality constraints.
(This inverse problem method is similar to the gradient-based forward method described in Ref.~\citenum{zhang_gradient_2014}.)
It does, however, suffer from the same rounding errors involved in the Wu-Yang algorithm as it still involves a minimization of the noninteracting kinetic energy.
Furthermore, the orthonormality and density constraints can easily lead to unphysical orbitals as they compete with one another in the optimization.

In order to resolve the issues mentioned above, we use the scaled orbitals introduced in \myref{sec:pde-constrained} and regularize them by adding the penalty functional
\begin{equation}
R\left(\left\{ g_{j}\right\} ,\alpha\right)=\alpha\sum_{j=1}^{N_{\text{orbs}}}\int\text{d}{\bf r}\left|\boldsymbol{{\bf \nabla}}g_{j}\!\left({\bf r}\right)\right|^{2},\label{eq:regularize-scaled-orb}
\end{equation}
to the cost functional $J\left(\left\{ \phi_{j}\right\} \right)$, where $\alpha>0$ is the regularization parameter chosen via the discrepancy principle.\cite{vogel_computational_2002}
The density constraint with the scaled orbitals is 
\begin{equation}
\int\text{d}{\bf r}\left[\sum_{j=1}^{N_{\text{orbs}}}g_{j}^{2}\!\left({\bf r}\right)\right]-1.\label{eq:scaled-nconstraint}
\end{equation}
No boundary conditions are applied to the orbitals for reasons explained in \myref{sec:pde-constrained}.

\begin{figure}[h]
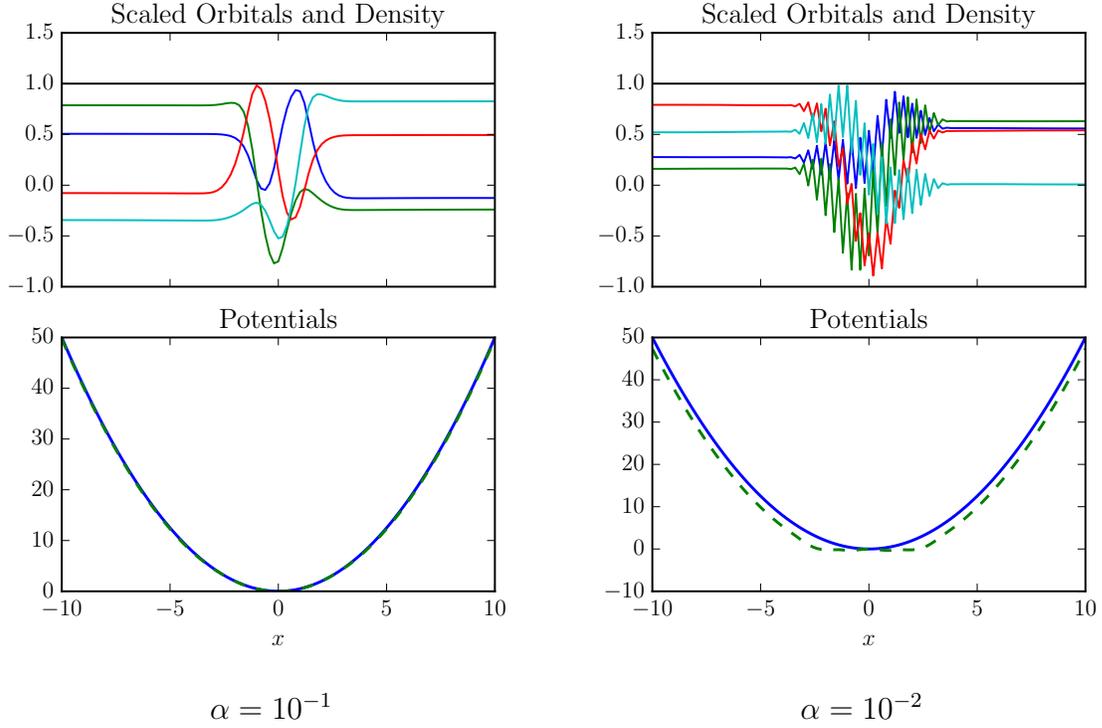

\begin{tabular}{cc}
\includegraphics[width=0.45\textwidth]{\myfigs{sho_alpha1e-1}} & \includegraphics[width=0.45\textwidth]{\myfigs{sho_alpha1e-2}}\tabularnewline
$\alpha=10^{-1}$ & $\alpha=10^{-2}$\tabularnewline
\end{tabular}
\caption[Constrained variational density-to-potential inversions with two different regularization parameters.]{\label{fig:sho_alpha}The scaled orbitals and density (top) of two constrained variational density-to-potential inversions with $\alpha=10^{-1}$ (left) and $\alpha=10^{-2}$ (right).  The scaled density is the flat black line with a value of 1 for all $x$.  The orthonormalization constraint dominates in the under-regularized example (right) producing wild oscillations in the high-density region. The corresponding potentials (bottom) are found by solving Eq.~\eqref{eq:levy-dot-product}.}
\end{figure}

Optimizing the noninteracting kinetic energy using the scaled and regularized orbitals produces a set of orthonormal orbitals that are a unitary transformation of the \gls{ks} orbitals.\cite{parr_density-functional_1994}
We use the \textsc{ipopt} library\cite{wachter_implementation_2005} to perform this constrained minimization.
The \gls{ks} potential can be extracted from these orbitals by solving the set of linear equations
\begin{multline}
\Biggl\{\int\text{d}{\bf r}\,\phi_{\beta}\!\left({\bf r}\right)\Delta\phi_{\alpha}\!\left({\bf r}\right)=2\int\text{d}{\bf r}\,\phi_{\beta}\!\left({\bf r}\right)\phi_{\alpha}\!\left({\bf r}\right)v_{\text{KS}}\!\left({\bf r}\right)\\
+\left(\sum_{m=1}^{\alpha}\varepsilon_{m,\alpha}\phi_{m}\!\left({\bf r}\right)+\sum_{n=\alpha}^{N_{\text{orbs}}}\varepsilon_{\alpha,n}\phi_{n}\!\left({\bf r}\right)\right)\Biggr\}\label{eq:levy-dot-product}
\end{multline}
where $\alpha,\beta=1,\ldots,N_{\text{orbs}}$.
This linear system can be solved efficiently using a sparse least-squares solver in which the full matrix is never explicitly created.
An example of this density-to-potential inversion procedure with two different regularization parameters is shown in Fig.~\ref{fig:sho_alpha} for a target density formed from the sum of the ground state and first two excited state densities of the harmonic oscillator.

\mysection{\sffamily \Large Inversion Examples}\label{sec:inversion-examples}
We have chosen to use well-known noninteracting quantum systems with which most readers are familiar to help illustrate the inversion methods described in \myref{sec:dft-inverse}.
The \gls{ks} potentials recovered for these systems are equal to the known external potential because the systems are noninteracting and this allows us to focus on the numerical properties of each inversion algorithm.
More realistic \gls{dft} inversions use interacting densities and extract the exchange-correlation potential from the recovered \gls{ks} potential using Eq.~\eqref{eq:ks-potential-inverse}.
Unless otherwise noted, the examples we present all use the one-orbital approximation as the initial guess for the unknown potential.
The gradients used for the \gls{pde}-constrained optimizations are given in App.~\ref{app:discrete-adjoint}.
Table~\ref{tab:ti-inverse-labels} displays the abbreviations used in this section to distinguish the different inversion methods.

In these examples we implement scaling, different finite difference approximations, and placement of the boundary conditions in the potential for both the PDE and CV routines.
We do this simply because we our more familiar with these methods and already have this capability implemented in our code.
It should also be possible to modify both the vLB and WY methods in similar ways to implement these features so that all of the schemes have similar error patterns in the asymptotic regions.
All of the error patterns in these examples are shifted relative to one another so as to see the differences and make better comparisons.

\begin{table}[h]
\caption{\label{tab:ti-inverse-labels}The labels used to identify each \gls{dft}
inversion method in \myref{sec:inversion-examples}.}

\centering{}%
\begin{tabular}{cc}
Inversion Method & Label\tabularnewline
\hline 
PDE-constrained optimization & PDE\tabularnewline
van Leeuwen and Baerends & vLB\tabularnewline
Wu and Yang & WY\tabularnewline
Constrained Variational & CV\tabularnewline
\end{tabular}
\end{table}

\begin{figure}[h]
\begin{center}
\includegraphics[width=0.65\textwidth]{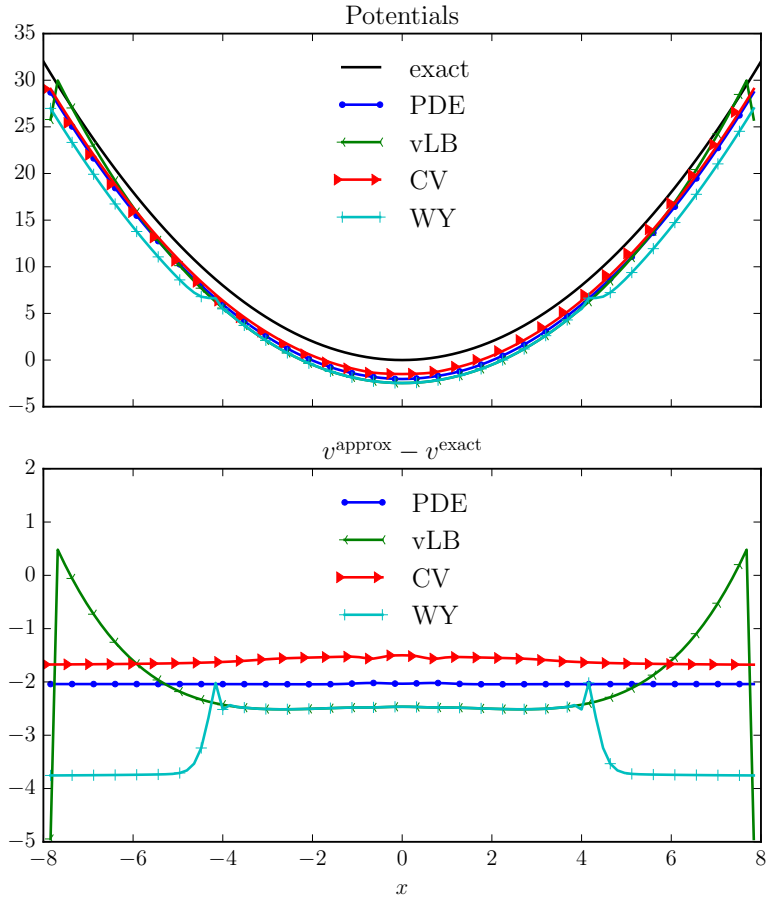}
\end{center}
\caption[A comparison of different density-to-potential inversion methods for a target density formed from the sum of the ground state and first two excited states of the harmonic oscillator.]{\label{fig:compare_sho} The potentials produced in a density-to-potential inversion using a target density formed from the sum of the ground state and first two excited states of the harmonic oscillator (top).  The correct potential is only unique up to a constant so we also plot the difference $v^{\mathrm{approx}}-v^{\mathrm{exact}}$ (bottom) with the more constant lines being an indication of correctness.}
\end{figure}

\mysubsection{\sffamily \large Harmonic Potential Inversion}\label{sec:hp-example}
Our first inversion example is a noninteracting system of six electrons in the harmonic potential $v\!\left(x\right)=\frac{1}{2}x^{2}$.
The target density is constructed from the lowest 3 orbitals
\begin{align}
\phi_{0}\!\left(x\right) & =\pi^{-1/4}e^{-\frac{x^{2}}{2}},\label{eq:hpt-phi0}\\
\phi_{1}\!\left(x\right) & =\frac{\sqrt{2}x}{\pi^{1/4}}e^{-\frac{x^{2}}{2}},\textrm{ and}\label{eq:hpt-phi1}\\
\mathbf{\phi}_{2}\!\left(x\right) & =\frac{\left(2x^{2}-1\right)}{\sqrt{2}\pi^{1/4}}e^{-\frac{x^{2}}{2}},\label{eq:hpt-phi2}
\end{align}
with each orbital doubly occupied as indicated in Eq.~\eqref{eq:ks-density-direct}.
The potentials recovered from this density using several different inversion methods are shown in Fig.~\ref{fig:compare_sho} using a grid of 101 equally-spaced points running from -8 to 8.
The potential is only unique up to a constant so we also plot the differences between the recovered and exact potentials with the more constant differences indicating better recovered potentials.

In this example, the rounding errors in the WY algorithm's functional dominate in the region $\left|x\right|>4$ as explained in App.~\ref{app:numerical-issues}.
The unscaled finite-difference scheme used in the vLB method is not accurate in the region $\left|x\right|>4$ and leads to an incorrect higher curvature in that region.
Both the CV and PDE methods have similar problems when they are not scaled or use zero boundary conditions so this example is not meant to show the superiority of any one scheme but rather to show the importance of these modifications.

\begin{figure}[h]
\begin{center}
\includegraphics[width=0.65\textwidth]{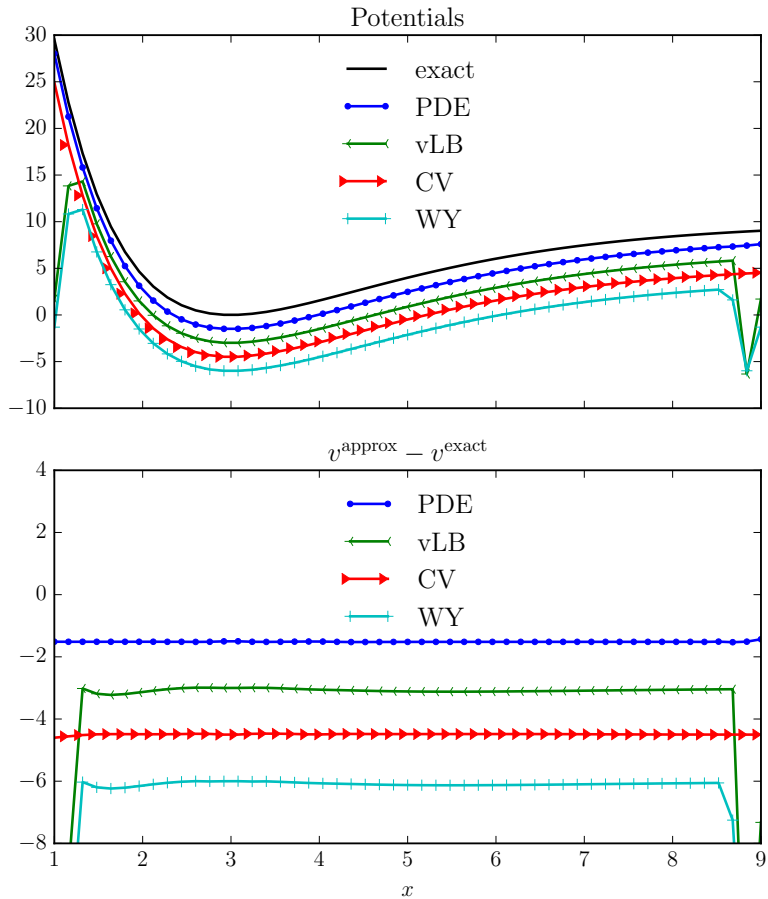}
\end{center}
\caption[A comparison of different density-to-potential inversion methods for a target density formed from the sum of the ground state and first two excited states of the Morse oscillator.]{\label{fig:compare_morse} The potentials produced in a density-to-potential inversion using a target density formed from the sum of the ground state and first two excited states of the Morse oscillator (top).  The correct potential is only unique up to a constant so we also plot the difference $v^{\mathrm{approx}}-v^{\mathrm{exact}}$ (bottom) with the more constant lines being an indication of correctness.  Both the PDE and CV methods are scaled and use fourth-order approximate Laplacians compared to the second-order schemes used in the vLB and WY methods.}
\end{figure}

\mysubsection{\sffamily \large Morse Potential Inversion}\label{sec:morse-example}
Our second inversion example is a noninteracting system of six electrons in the Morse potential
\begin{equation}
v\!\left(r\right)=D_{e}\left(1-e^{-a\left(r-r_{e}\right)}\right),\label{eq:morse-potential}
\end{equation}
where $a=1/2$, $D_{e}=10$, and $r_{e}=3$.
Analytic solutions for this system's wave functions can be found in Ref.~\citenum{dahl_morse_1988} and are used as input in Eq.~\eqref{eq:ks-density-direct} for forming the target density.
The potentials recovered from the target density described above are shown in Fig.~\ref{fig:compare_morse} using a grid of 51 equally-spaced points running from 1 to 9.

This inversion example illustrates some of the numerical issues described in \myref{sec:dft-theory}.
One numerical issue is the need to adequately represent the kinetic energy operator on the chosen grid.
The PDE and CV methods in this example use a fourth order approximation of the Laplacian while the other methods use a second-order approximation.
This difference in accuracy on such a coarse grid leads to several wiggles in the WY and vLB potentials because the inversion routines fit to the errors in the approximate Laplacian.
Additionally, the one-orbital approximation is not a sufficiently accurate initial guess for the scaled PDE-constrained optimization to locate the correct potential, hence the optimizer gets stuck in a local minimum.
(The potential is so shallow that the scaled density is much too large near the right boundary.)
This can be remedied by either raising the boundary values of the initial guess to force a more rapid decay or by including a small amount of regularization via Eq.~\eqref{eq:regularize-scaled-orb} with $\alpha=1$ to guide the initial optimization and then turning it off with $\alpha=0$ before running the final optimization.
The PDE result shown in Fig.~\ref{fig:compare_morse} uses the regularization option but the result is nearly identical to that found when raising the boundary values of the initial guess.
Both of these issues arise even when we have an analytic target density with no added error.

\mysubsection{\sffamily \large Kronig-Penney Inversion}\label{sec:kronig-example}
Our last inversion example is a noninteracting system of six electrons per unit cell in the Kronig-Penney potential
\begin{equation}
v\!\left(x\right)=\begin{cases}
V_{0} & \text{if }mL-\frac{a}{2}\le x\le mL+\frac{a}{2}\\
0 & \text{otherwise}
\end{cases},\label{eq:kronig-penney}
\end{equation}
where $L=2$ is the length of the unit cell, $a=5/4$ is the width of the well, $V_{0}=-10$ is the depth of the well, and $m$ is an integer.
The target density is found by extracting the density from the center cell of a finite system comprised of 15 unit cells, [$m=-7,\ldots,7$ in Eq.~\eqref{eq:kronig-penney}], with 400 points per unit cell and a second-order discretization of the Laplacian.
By performing a finite calculation of the target density we avoid the risk of committing an inverse crime.

\begin{figure}[h]
\begin{center}
\includegraphics[width=0.65\textwidth]{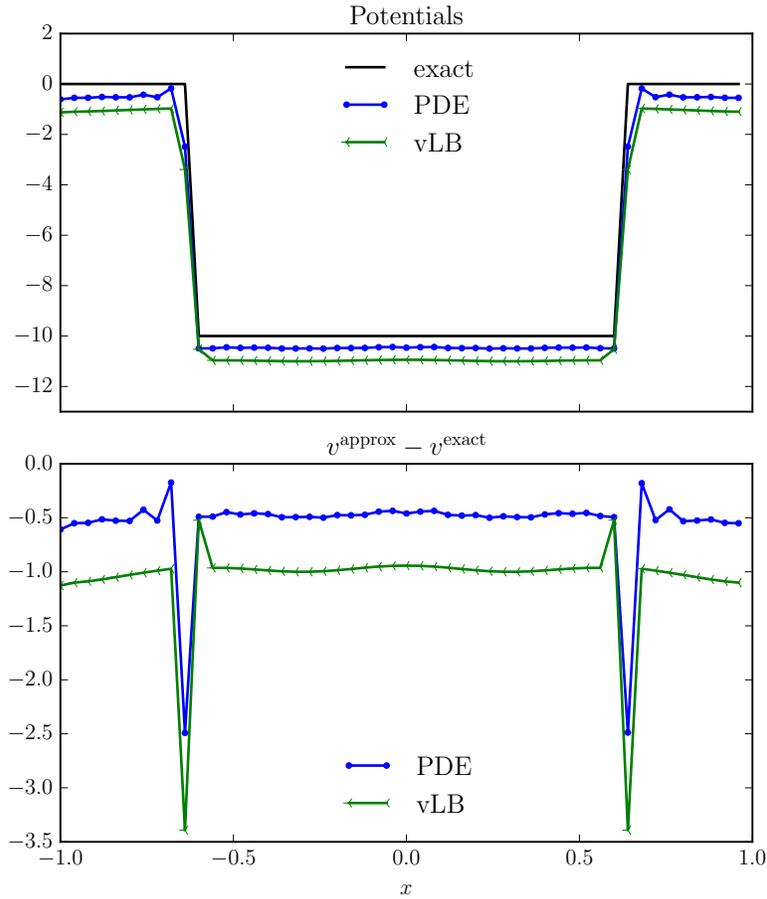}
\end{center}
\caption[A comparison of different density-to-potential inversion methods for a target density formed from the sum of the ground state and first two excited states of the Kronig-Penney periodic system.]{\label{fig:compare_kronig} The potentials produced in a density-to-potential inversion using a target density formed from the sum of the ground state and first two excited states of the Kronig-Penney periodic system (top).  The difference $v^{\mathrm{approx}}-v^{\mathrm{exact}}$ (bottom) shows that most of the error is near the jumps at $-a/2$ and $a/2$.}
\end{figure}

The inverse problem for this example is solved on a grid with 50 points using \gls{pde}-constrained optimization without scaling and a fourth-order approximate Laplacian.
(We don't use scaling because the exponential decay doesn't dominate anywhere in the system's density.)
The Hamiltonian is modified as explained in Ref.~\citenum{castro_octopus:_2006} to work with the Bloch functions $\left\{ \phi_{i,{\bf k}}\!\left({\bf r}\right)\right\} $.
In particular, the kinetic energy operator has a $k$-space dependence
\begin{equation}
\hat{T}_{{\bf k}}=-\frac{1}{2}\left(\Delta+2\imath{\bf k}\cdot\boldsymbol{\nabla}-k^{2}\right)\label{eq:kinetic-periodic-op}
\end{equation}
and the total density is found via an integration over the first Brillouin zone
\begin{equation}
n\!\left({\bf r}\right)=\frac{1}{\Omega_{{\bf k}}}\int_{\Omega_{{\bf k}}}\text{d}{\bf k}\sum_{i=1}^{N_{\text{orbs}}}\left|\phi_{i,{\bf k}}\!\left({\bf r}\right)\right|^{2},\label{eq:density-integration-brillouin}
\end{equation}
where $\Omega_{{\bf k}}$ is the volume of the first Brillouin zone.
We perform this $k$-space integration using the trapezoid rule and 11 equally spaced points running from $-\pi/L$ to $\pi/L$.
(For simplicity we integrate over the whole first Brillouin zone but usually symmetry is used to reduce this to an integral over the irreducible Brillouin zone.)
The results of this inversion are shown in Fig.~\ref{fig:compare_kronig} and show that even sharp features such as Heaviside functions can be recovered successfully in these inversions.
\section*{\sffamily \Large Conclusions}

We have presented a series of algorithms for performing density-to-potential inversions in \gls{dft}.
Although the examples in this work are all one-dimensional model systems, the algorithms are written in multiple dimensions and are applicable to more realistic systems.
In particular, these methods can all be used to find the exact exchange-correlation potential corresponding to a given interacting density as explained in Sec.~\ref{sec:dft-theory}.
We have also presented a detailed numerical analysis that points to the relative error as the main limitation in density-to-potential inversions and the need for error estimates in performing these inversions properly.

The inversion methods presented in this tutorial use the finite difference method for solving the governing \glspl{pde} but they can also be applied to the common basis function methods described in Ref.~\citenum{martin_electronic_2004}.
The insight we have gained through our many test cases and numerical analysis with the finite difference method can also be used to improve other inversion methods as hinted at in \myref{sec:hp-example}.
In particular, the scaled-orbital regularization presented in \myref{sec:var-constrained} is perhaps a more natural regularization than the potential regularization presented in Ref.~\citenum{bulat_optimized_2007}.
This regularization should also make it possible to examine more realistic examples including possibly ill-posed densities from experimental data.

All of the inversion methods used in Sec.~\ref{sec:inversion-examples} have strengths and weaknesses that should be considered when choosing which method to use for a given inverse \gls{dft} problem.
The PDE-constrained optimization algorithm is very much a black-box method requiring almost no physical intuition but it is only competetive with other methods when the gradient is computed efficiently and used in a good optimization routine.
This is not a trivial endeavor as both the derivation and programming of these gradients can be quite involved as shown in App.~\ref{app:discrete-adjoint}.
Despite this weakness, it is very straightforward to adapt this method to different inverse problems resulting from extensions of \gls{dft} such as \gls{tddft}.\cite{jensen_numerical_2016}
The method of van Leeuwen and Baerends does have a sound physical motivation and is simple to program because it only needs to solve the Kohn-Sham equations.
The biggest disadvantage to this method is its reliance on a prefactor, which requires user intervention and makes automating the routine difficult.
The method of Wu and Yang is similar to the van Leeuwen and Baerends method in that it is also based on some physical motivation and only needs to solve the Kohn-Sham equations provided the Hessian isn't needed.
This method is very efficient when paired with a good optimizer but can suffer from rounding and overfitting errors if extra precautions are not taken.
The constrained variational method is the only method that doesn't solve an eigenvalue problem at each iteration and this may be an advantage for large systems with more demanding eigenvalue problems.
In the systems that we have studied this method appears to be insensitive to the initial guess with even random numbers for the initial orbitals resulting in rapid convergence.
Its largest drawback is its current reliance on a regularization parameter but this choice can be automated using standard inverse problem methods.
Since we have only considered model one-dimensional systems in this study we do not know how these methods will perform on more realistic three-dimensional systems and plan to do future studies to see how they scale with larger systems.

Our original intent in developing inversion algorithms was to benchmark approximate exchange-correlation approximations to guide new approximations within \gls{dft}.
During this development, however, several new and interesting applications have come to light that warrant further study.
For example, by allowing the inversion methods to enforce the boundary conditions by modifying the potential near the boundaries in the \gls{dcbc} method, it should be possible to extract new approximate boundary conditions and compare them to other state-of-the-art methods.\cite{antoine_review_2008}
Likewise, the orbital scaling introduced in \myref{sec:pde-constrained} could be modified for the direct \gls{dft} problem as an alternative to using adaptive grid methods.\cite{castro_octopus:_2006}
Thanks to the very similar structure between the \gls{oep} method and the \gls{dft} inverse problem, algorithm development in one area can lead directly to improvements in the other as seen in the development of the Wu-Yang inversion algorithm from a previous \gls{oep} optimization method.\cite{wu_direct_2003}
We plan to continue exploring these other areas of \gls{dft} and refine the density-to-potential inversion methods described in this tutorial.

%
%
%
%
%
%
%
%
%
%
%

\subsection*{\sffamily \large Acknowledgments}

Sandia National Laboratories is a multimission laboratory managed and operated by National Technology and Engineering Solutions of Sandia, LLC., a wholly owned subsidiary of Honeywell International, Inc., for the U.S. Department of Energy’s National Nuclear Security Administration under contract DE-NA0003525.
This work was supported by the National Science Foundation CAREER program under Grant No. CHE- 1149968.
A.W. also acknowledges support from the Camille Dreyfus Teacher-Scholar Awards Program.
We thank Jonathan Nafziger for many discussions related to \gls{dft} inversions.
We also thank Andrew Baczewski, Attila Cangi, and the referees for their many helpful comments.

\begin{appendices}
\section{\sffamily \Large Numerical Issues}\label{app:numerical-issues}
A number of numerical problems arise in performing \gls{dft} inversions that are often very difficult to isolate.
In this appendix we use very basic one-dimensional models to show how boundary conditions, relative error, and rounding errors all set convergence limits on density-to-potential inversions.
We employ the one-orbital formula shown in Eq.~\eqref{eq:one-orbital-ti} and a grid ranging from $-4$ to $4$ with 51 points in the examples of this section unless otherwise noted.

\subsection{\sffamily \large Boundary Conditions and Inverse Crimes}\label{app:boundary-conditions}
\begin{figure}[h]
 \centering
 \includegraphics[width=.4\textheight]{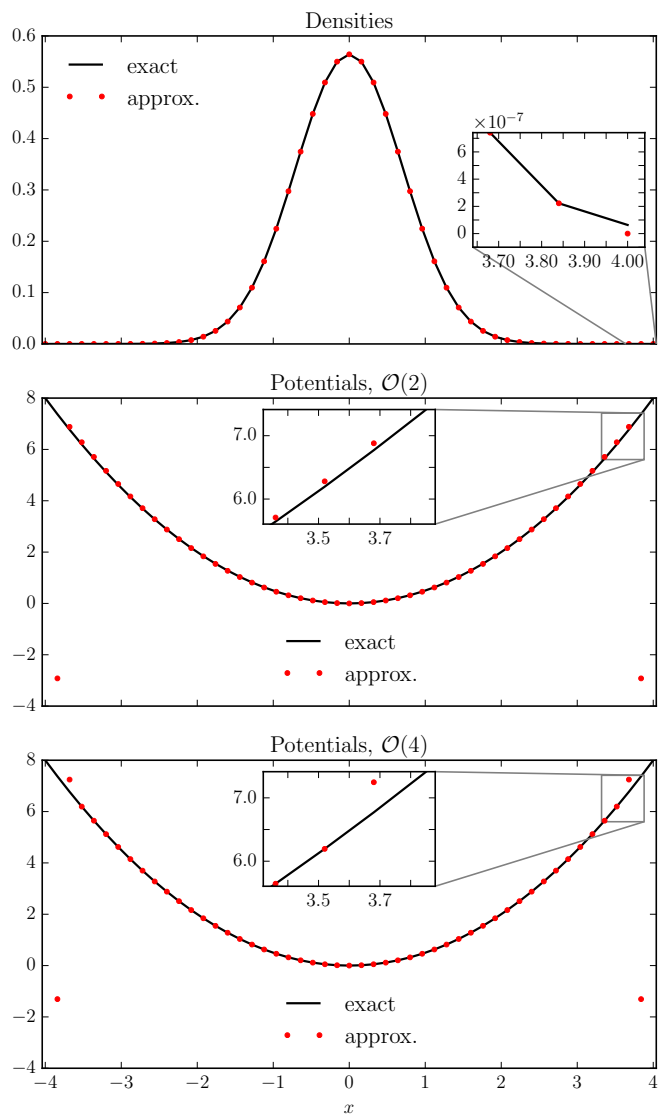}
 \caption[The degradation of a density-to-potential inversion when forcing incorrect boundary conditions on the orbitals.]{The degradation of a density-to-potential inversion when forcing incorrect boundary conditions on the orbitals.  The exact ground-state density of the simple harmonic oscillator is modified to be exactly zero at the boundaries as commonly occurs when box-type (Dirichlet) boundary conditions are enforced (top).  The potentials produced from this density with the one-orbital inversion formula and a second-order discretization of the Laplacian are incorrect near the boundaries (middle).  (The outermost points have the largest error and we zoom in on the next closest values to show their errors.)  The errors are even larger when a fourth-order approximation to the Laplacian is used (bottom).}
 \label{fig:mismatch_boundaries}
\end{figure}

Many of the inversion schemes mentioned in \myref{sec:dft-inverse} impose boundary conditions on the orbitals that don't have to agree with the target density's boundary conditions.
A mismatch in boundary conditions can place severe constraints on the accuracy of density-to-potential inversions in the boundary regions as we illustrate here with an example.
For simplicity, we take the exact density of the simple harmonic oscillator's ground state and set the values at the boundaries to zero as would be typical for box-type (Dirichlet) boundary conditions.
The one-orbital formula should give us the exact potential for this problem everywhere except the boundaries, (where the density was artificially set to zero), but Fig.~\ref{fig:mismatch_boundaries} shows that the potential is also incorrect at several points next to the boundaries.
The number of incorrect points near the boundaries increases as higher-order finite difference schemes are used to represent the Laplacian due to the increased nonlocality of the finite difference operators.

\begin{figure}[h]
 \centering
 \includegraphics[width=.6\textwidth]{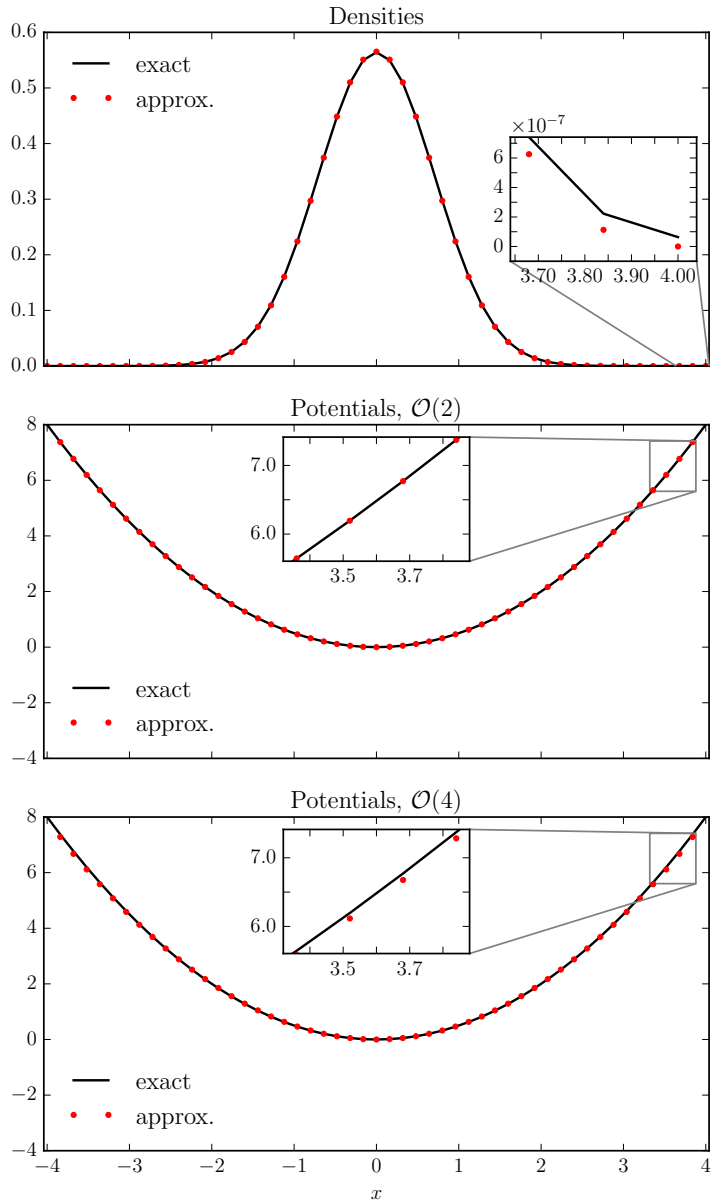}
 \caption[The frailty of inverse crimes in a density-to-potential inversion.]{The frailty of inverse crimes in a density-to-potential inversion.  The ground state density of the harmonic oscillator is computed numerically using a second-order approximation to the Laplacian and zero boundary conditions (top, approx).  Applying the one-orbital formula to this density with a second-order approximate Laplacian produces nearly the exact potential (middle) and is an inverse crime. A fourth-order approximation in the one-orbital formula actually gives worse values most noticeable near the boundaries (bottom).}
 \label{fig:inverse_crime}
\end{figure}

One common method for avoiding the mismatch in boundary conditions described above is to compute the target density with the same boundary conditions and numerical method.
In Fig.~\ref{fig:inverse_crime} we compute the ground state density of the harmonic oscillator numerically with box-type boundary conditions and a second-order finite difference approximation for the Laplacian.
If the same finite difference approximation is used in the one-orbital formula then the resulting potential is almost exact everywhere.
This type of inversion is known as an inverse crime.
It is easily seen in Fig.~\ref{fig:inverse_crime} that changing to a fourth-order approximate Laplacian with the same box-type boundary conditions does not have the same error cancellation and produces worse values than the second-order method.

The term `inverse crime' is commonly used in inverse problem theory to describe an inversion method relying on a cancellation of errors between the data simulation and reconstruction methods.\cite{muller_linear_2012}
Inverse crimes are only committed in \gls{dft} inversions when the same numerical methods are used to solve both the direct and inverse problems.
This is somewhat rare in performing \gls{dft} inversions because the densities being inverted are usually found using wave-function methods that employ different numerical procedures than those used in \gls{dft} inversions.\cite{baker_one-dimensional_2015}
Inversion methods that rely on inverse crimes are very limiting in the number of systems that can be studied and will likely fail when applied to experimental data.

This example of an inverse crime is not meant to discourage practitioners from using knowledge about the target density in performing inverse \gls{dft} problems.
Much of Sec.~\ref{sec:one-orb} is devoted to using this knowledge to better approximate the true density and invert it to find the \gls{ks} potential.
An alternative is to use similar boundary conditions and numerical methods for both the interacting system and inverse problem with the understanding that there will be some cancellation of errors if the methods are similar enough.
This is not really an inverse crime since the solution method for solving the interacting system is not identical to the numerical methods for solving the \gls{ks} equations but it is not a requirement when solving these problems.
In fact, the methods presented in Sec.~\ref{sec:dft-inverse} can be applied to target densities resulting from a variety of different numerical methods with no need to rely on error cancellations.

\subsection{\sffamily \large Target Densities with Numerical Noise}\label{app:target-density-noisy}
\begin{figure}[h]
 \centering
 \includegraphics[width=.6\textwidth]{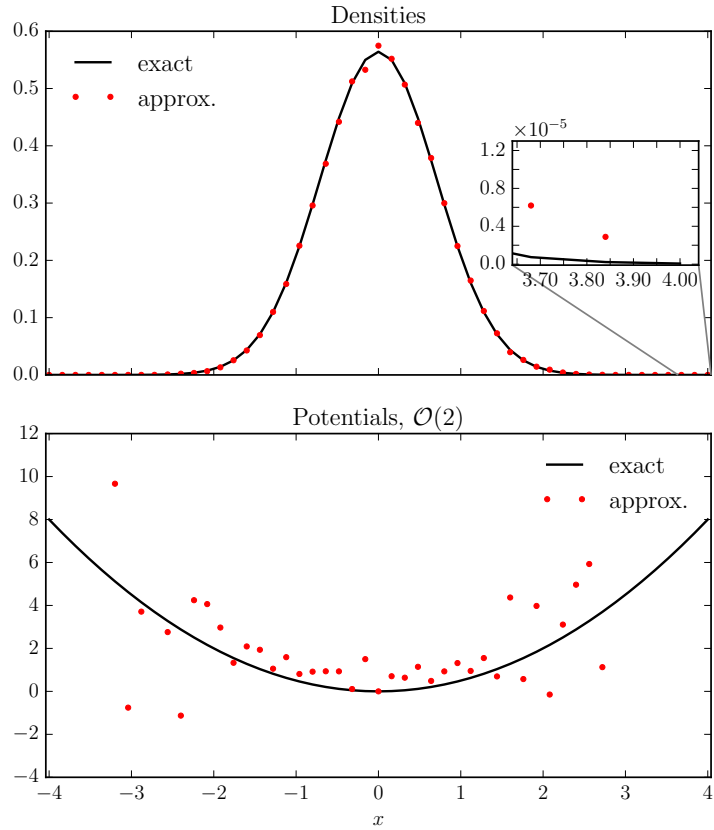}
 \caption[The effects of random noise in a density-to-potential inversion.]{The effects of random noise in a density-to-potential inversion.  When the ground-state of the simple harmonic oscillator contains noise, (i.e. $\tilde{n}_{0}=\left[\phi_{0}+\ensuremath{N\!\left(\mu=0,\sigma^{2}=2.5\times10^{-5}\right)}\right]^{2}$), the resulting potential is very noisy everywhere but especially in the asymptotic region.  The gauge in this example is fixed by forcing the value of the inverted potential to agree at $x=0$ with the exact potential.}
 \label{fig:phi_err51}
\end{figure}

\begin{figure}[h]
 \centering
 \includegraphics[width=.6\textwidth]{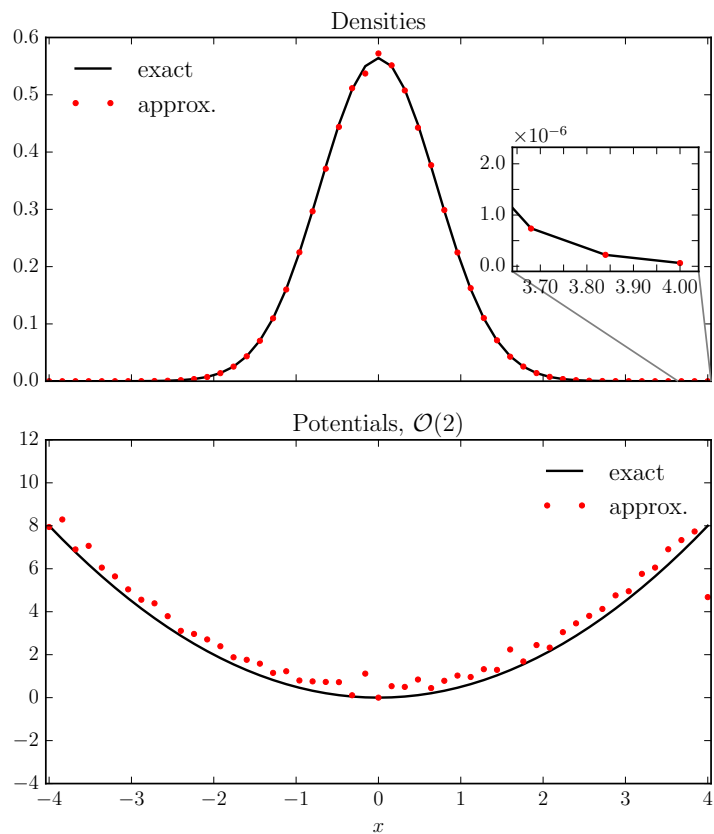}
 \caption[The effects of weighted random noise in a density-to-potential inversion.]{The effects of weighted random noise in a density-to-potential inversion.  When the ground-state of the simple harmonic oscillator contains noise weighted by the orbital, (i.e. 
 $\tilde{n}_{0}=\left\{ \phi_{0}\left[1+\ensuremath{N\!\left(\mu=0,\sigma^{2}=2.5\times10^{-5}\right)}\right]\right\} ^{2}$), the resulting potential is uniformly noisy.  This indicates that the relative error in the target density sets a convergence limit in density-to-potential inversions.  The gauge in this example is fixed by forcing the value of the inverted potential to agree at $x=0$ with the exact potential.}
 \label{fig:phi_relerr51}
\end{figure}

\begin{figure}[h]
 \centering
 \includegraphics[width=.6\textwidth]{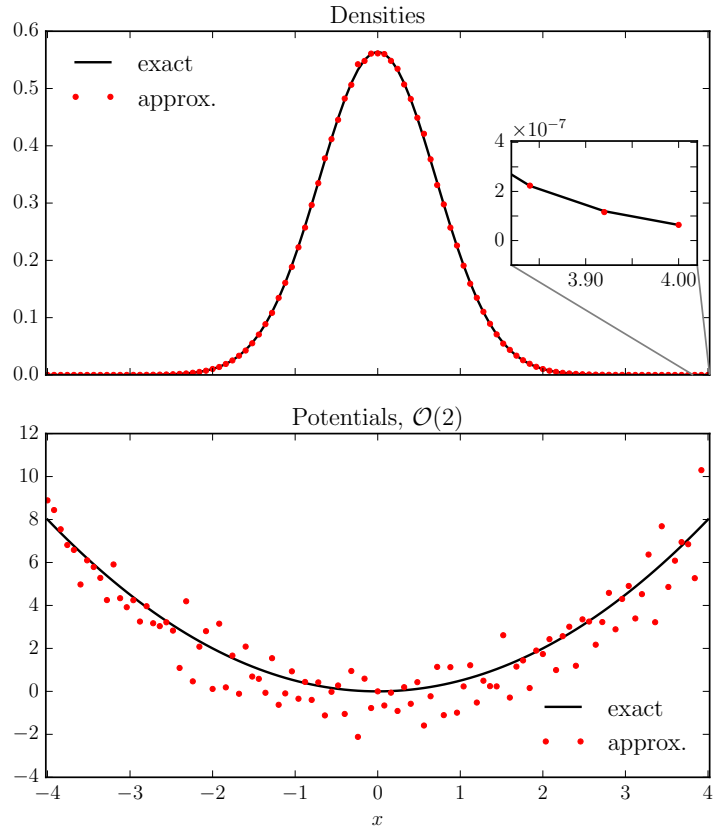}
 \caption[The effects of weighted random noise and increased grid resolution in a density-to-potential inversion.]{The effects of weighted random noise and increased grid resolution in a density-to-potential inversion.  When twice the number of grid points are used for the example displayed in Fig.~\ref{fig:phi_relerr51} the resulting potential is actually worse and shows that increased resolution does not necessarily improve density-to-potential inversions.  The gauge in this example is fixed by forcing the value of the inverted potential to agree at $x=0$ with the exact potential.}
 \label{fig:phi_relerr101}
\end{figure}

The examples in the preceding section assume a perfect knowledge of the target density but this is not a realistic assumption for most inverse problems.
As mentioned in the previous example, the wave function numerical methods used in computing interacting densities are often very different from \gls{dft} numerical methods and the correct boundary conditions are not always known beforehand or may even be incorrect.
It is far more realistic to assume that the target density in a density-to-potential inversion has some numerical error that must be accounted for to avoid overfitting.
In order to illustrate the effects of numerical noise in the target density, we take the ground state orbital of the simple harmonic oscillator, add in a small amount of noise generated from the standard normal distribution 
$\ensuremath{\operatorname{N}\!\left(\mu,\sigma^{2}\right)}$,
\cite{bain_introduction_2000} and put the density corresponding to this modified orbital in the one-orbital formula.
The results of this inversion in Fig.~\ref{fig:phi_err51} are very poor especially in the asymptotic region.
If we add a relative error to the ground state orbital instead of the uniform error described above then the resulting potential is still noisy but the noise level is now the same everywhere as seen in Fig.~\ref{fig:phi_relerr51}.
These two examples show that the relative error of the target density sets a limit on the convergence in density-to-potential inversions.
Figure \ref{fig:phi_relerr101} shows that simply increasing the number of grid points actually makes the inverted potential worse when the same amount of error is present in the target density.

\subsection{\sffamily \large Finite Difference Operators}\label{app:finite-diff-operators}
\begin{figure}[h]
 \centering
 \includegraphics[width=.6\textwidth]{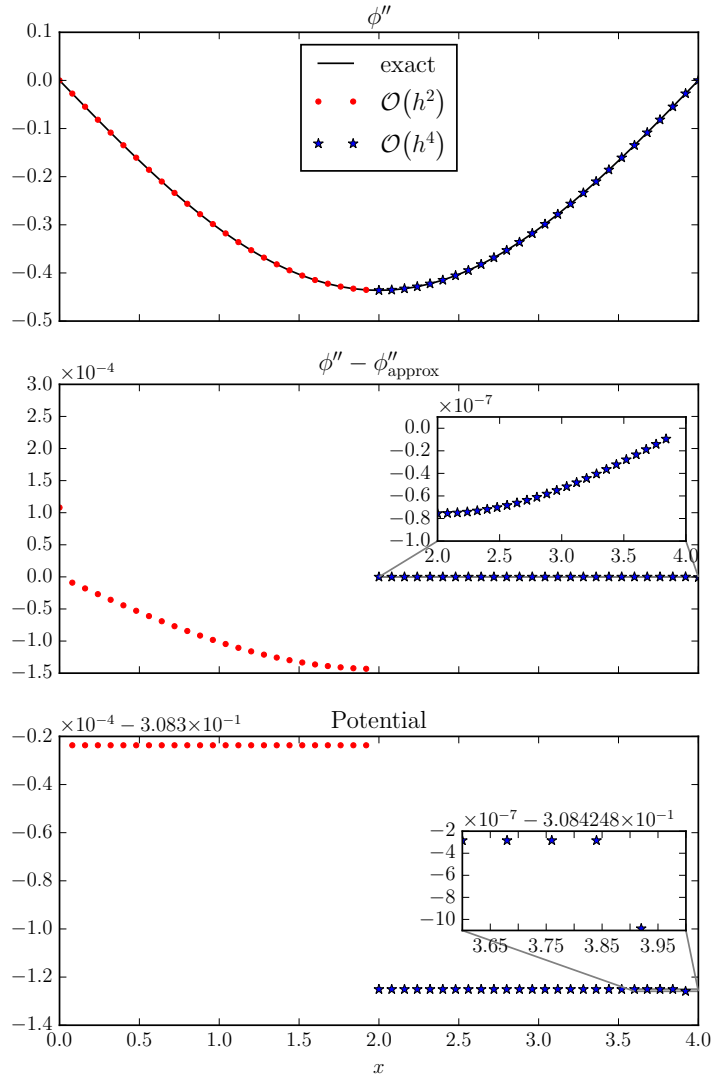}
 \caption[The effects of combining two different finite difference operators in a density-to-potential inversion.]{The effects of combining two different finite difference operators in a density-to-potential inversion.  
 The second derivative is computed with a second-order method on the left side of the box and a fourth-order method on the right (top).
 The error patterns on both sides of the box are smooth but there is a distinct gap between them (middle).
 The resulting potential (bottom) also has a gap corresponding to the difference in error patterns and a small gap at the right boundary where a backwards finite-difference operator is used (bottom inset).  The scale on the y-axis of the lower two plots is actually very small ($10^{-4}$) so these errors are not usually seen but they do make a difference in the inverse problem.}
 \label{fig:mismatch_ops}
\end{figure}

As mentioned in the previous examples, higher-order finite difference operators don't always improve density-to-potential inversions and boundary conditions often have a dramatic impact on the quality of an inversion.
Another way to study the delicate balance between finite difference operators and boundary conditions is to use the ground state density of a particle in a box in the one-orbital formula.
If we compute the derivative in Eq.~\eqref{eq:one-orbital-ti} with one finite difference operator on the left side of the simulation box and a different operator on the right then the resulting second derivative has different error patterns on both sides of the box.
The resulting potential has a jump between the left and right sides of the box as seen in Fig.~\ref{fig:mismatch_ops}.
This jump is actually very small, (on the order of $10^{-4}$ as seen in the scale), but makes a significant impact on the error pattern.
Although one usually doesn't mix finite difference operators inside the simulation box, different operators are often used at the boundaries and produce similar jumps even if the operators are the same order of accuracy.
An example of this jump at the boundary can be seen in Fig.~\ref{fig:mismatch_ops} at the right boundary where a fourth-order backwards finite-difference operator produces a different error pattern than the fourth-order centered operator used elsewhere.
(See Eq.~\eqref{eq:fd-stencil-O4} for the matrix representation of this fourth-order finite-difference operator.)

\begin{figure}[h]
 \centering
 \includegraphics[width=.6\textwidth]{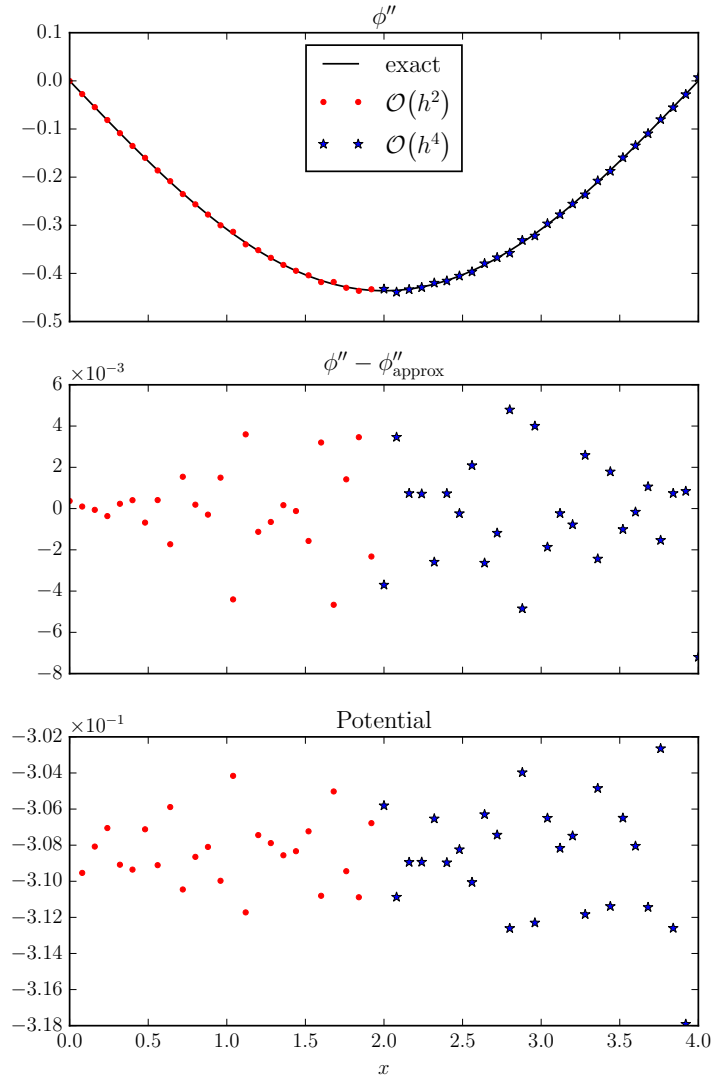}
 \caption[The effects of combining two different finite difference operators in a density-to-potential inversion with a noisy density.]{The effects of combining two different finite difference operators in a density-to-potential inversion with a noisy density.
 This example is identical to the one shown in Fig.~\ref{fig:mismatch_ops} except the input density contains noise, (i.e. $\tilde{n}_{0}=\left\{ \phi_{0}\left[1+\ensuremath{N\!\left(\mu=0,\sigma^{2}=1\times10^{-10}\right)}\right]\right\} ^{2}$).
 This is another example where higher-order methods can actually produce worse results than simpler lower-order methods.  The scales on the y-axis of the middle and bottom plots are $10^{-3}$ and $10^{-1}$ respectively.}
 \label{fig:mismatch_opserr}
\end{figure}

Many of the somewhat counterintuitive behaviors shown in the preceding examples can be understood mathematically by studying the effects of errors in the one-orbital formula.
If we have the exact density of a one orbital system then the potential is given by the formula
\begin{align}
v_{i} & =\frac{\phi_{i}''+\varepsilon_{i}^{\text{fd}}}{2\phi_{i}},\label{eq:one-orb-noise}
\end{align}
where $\phi_{i}=\sqrt{n_{i}}$ and $\varepsilon_{i}^{\text{fd}}$
is the error of the approximate second derivative at the point $x_{i}$.
In the example shown in Fig.~\ref{fig:mismatch_ops} there are different error patterns $\varepsilon_{i}^{\text{fd}}$ for the left- and right-hand sides of the box and this difference is magnified during the inversion because of the division by the square root of the density.
A more realistic example includes some numerical error $\varepsilon_{i}^{n}$ in the density being inverted and the inversion formula becomes
\begin{align}
v_{i} & =\frac{\left(\phi_{i}+\varepsilon_{i}^{n}\right)''}{2\left(\phi_{i}+\varepsilon_{i}^{n}\right)}=\frac{\phi_{i}''+\varepsilon_{i}^{\text{fd}}+\left(\varepsilon_{i}^{n}\right)''}{2\left(\phi_{i}+\varepsilon_{i}^{n}\right)}.\label{eq:one-orb-double-noise}
\end{align}
If the noise $\varepsilon_{i}^{n}$ is not smooth then its second derivative will be very noisy and dominate the other terms in Eq.~\eqref{eq:one-orb-double-noise}.
Figure~\ref{fig:mismatch_opserr} shows an example where a very small normally distributed error severely limits the accuracy of the potential.

\subsection{\sffamily \large Rounding Errors}\label{app:rounding-errors}
\begin{figure}[h]
 \centering
 \includegraphics[width=.6\textwidth]{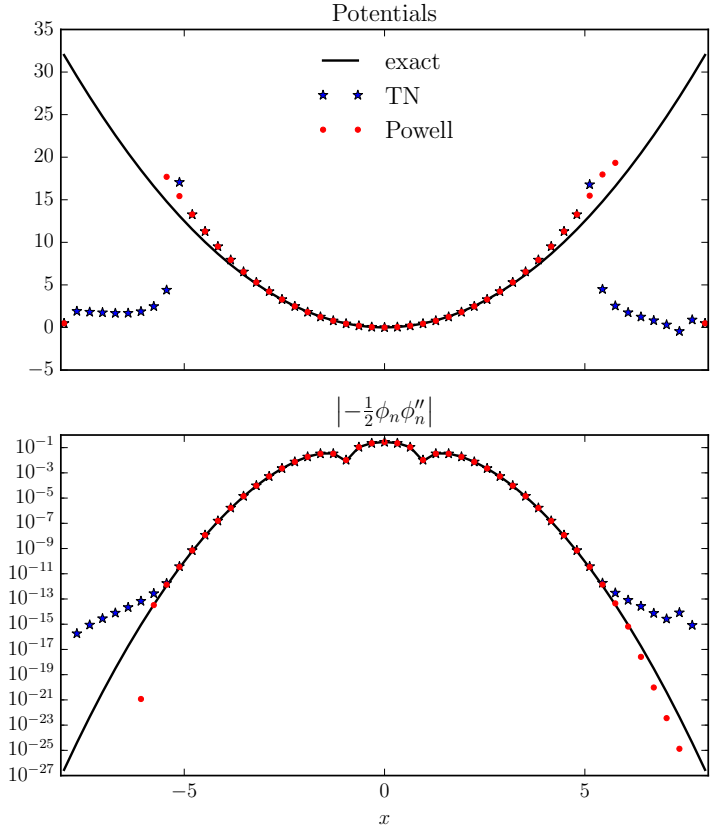}
 \caption[The effects of rounding errors in a density-to-potential inversion with double-precision numbers.]{The effects of rounding errors in a density-to-potential inversion with double-precision numbers.
 Both the \gls{tn} and modified Powell's optimizers fail to resolve the potential in the region $\left|x\right|>5.5$ (top) when used in the Wu-Yang algorithm and an initial guess of zero.
 The kinetic energy density is orders of magnitude smaller in the asymptotic region than it is in the region $\left|x\right|<5.5$ (bottom) and this leads to loss of precision when computing the total kinetic energy.}
 \label{fig:ke_cutoff}
\end{figure}

\begin{figure}[h]
 \centering
 \includegraphics[width=.6\textwidth]{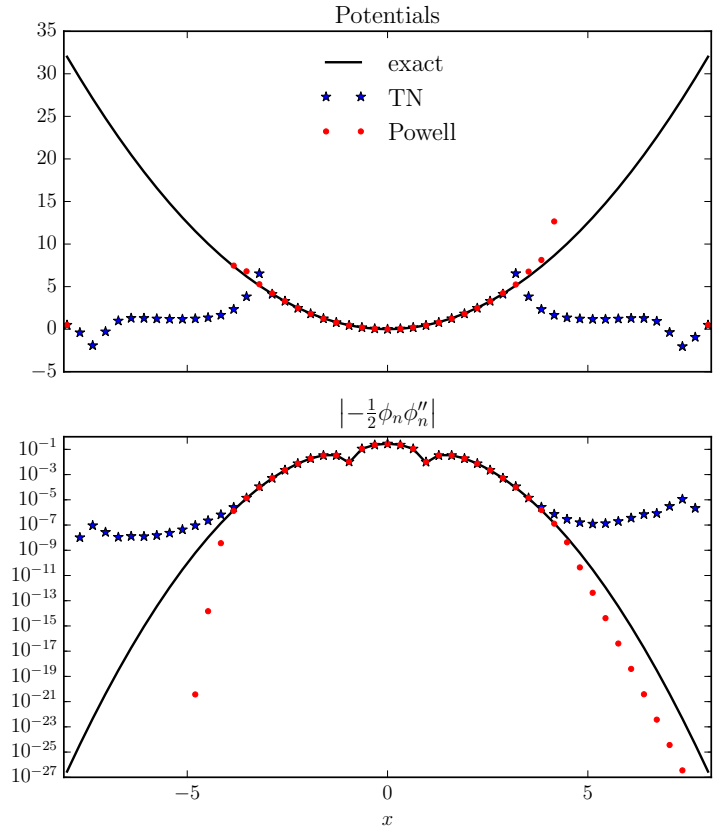}
 \caption[The effects of rounding errors in a density-to-potential inversion with single-precision numbers.]{The effects of rounding errors in a density-to-potential inversion with single-precision numbers.
 This example is identical to the one shown in Fig.~\ref{fig:ke_cutoff} except that the kinetic energy sum uses single-precision numbers.
 In this case the potential is only recovered in the region $\left|x\right|<3.5$ due to more severe rounding errors from the use of single-precision numbers.}
 \label{fig:ke_cutoff32}
\end{figure}

Rounding errors play an important role in many of the inversions methods dependent upon numerical optimization methods.
The Wu-Yang algorithm discussed in \myref{sec:dft-inverse} needs to compute the kinetic energy at each iteration of a given numerical optimization routine.
If the density of a system decays exponentially then the kinetic energy from the asymptotic regions is very small in comparison to values near the nuclei.
Even if the values of the kinetic energy density can be computed accurately in all regions of a simulation, adding all of the values together to get the total kinetic energy will often result in rounding errors due to the large differences in magnitude involved.
In Fig.~\ref{fig:ke_cutoff} we apply the Wu-Yang inversion algorithm to the ground-state density of the harmonic oscillator on a grid with 51 points ranging from $-8$ to 8.
Both the gradient-based \gls{tn} algorithm\cite{nash_newton-type_1984} and the non-gradient-based modified Powell's method\cite{powell_efficient_1964} fail to resolve the density in the asymptotic region, ($\left|x\right|>5.5$), when used in the Wu-Yang inversion method.
The abrupt cutoffs near $-5.5$ and 5.5 are a result of rounding errors when computing the kinetic energy with double-precision floating point numbers.
If we use the midpoint integration rule to compute the kinetic energy in the region $\left|x\right|<5.5$ the result in double precision arithmetic is 0.2500000000000231 and the same method used in the region $\left|x\right|>5.5$ produces the value $-2.32460041955568\times10^{-14}$.
When the two numbers above are added together, most of the smaller number's significant digits are lost and this loss of information limits the accuracy of the recovered potential in the asymptotic region.
The same example performed using single-precision floating point numbers gives an even stricter cutoff near $-3.5$ and 3.5 as seen in Fig.~\ref{fig:ke_cutoff32}.
In this second example the kinetic energy in the region $\left|x\right|<3.5$ is 0.25000963 when using single-point precision numbers compared to a value of $-9.6271506\times10^{-6}$ in the region $\left|x\right|>3.5$, which again results in a loss of significant figures when added together.

The preceding examples of time-independent density-to-potential inversions illustrate many of the pitfalls in performing numerical density-to-potential inversions.
Simply refining the grid or increasing the accuracy of the kinetic energy operator does not guarantee an improved density-to-potential inversion and in many cases will worsen the solution.
In cases where the error pattern is not known explicitly it is actually not correct to fit the density exactly as even small density errors can lead to large artifacts in the potential.
Finally, inversions reliant on numerical optimization routines often suffer from rounding errors, which set yet another hard limit on the convergence of density-to-potential inversions.
\section{\sffamily \Large Discrete-Adjoint Method}\label{app:discrete-adjoint}
In this appendix we derive the discrete adjoint equations for the \gls{pde}-constrained optimization method presented in Sec.~\ref{sec:pde-constrained}.
The discrete-adjoint method is a very efficient method for computing derivatives (discrete functional derivatives) in \gls{pde}-constrained problems.\cite{j._e._hicken_pde-constrained_2014}
Computationally, these derivatives amount to a sparse linear solve for each occupied orbital and are about the same computational cost as computing the cost functional.
The derivations give here are only valid for systems without degeneracies because this method is based on the algebraic method described in Ref.~\citenum{rudisill_numerical_1975}.
A similar derivation can be performed for systems with degeneracies using the method of Ref.~\citenum{van_der_aa_computation_2007}.

\begin{table}
\caption{\label{tab:adjoint-notation}The notation used in the discrete-adjoint
derivations.}

\centering{}%
\begin{tabular}{cl}
Notation & Description\tabularnewline
\hline 
${\bf i}$ & Discrete spatial indices $i_{x},i_{y},i_{z}$\tabularnewline
$\sum_{{\bf i}={\bf 1}}^{N_{{\bf x}}}$ & Sum over all spatial indices $\sum_{i_{x}=1}^{N_{x}}\sum_{i_{y}=1}^{N_{y}}\sum_{i_{z}=1}^{N_{z}}$\tabularnewline
$F_{n}$ & Cost functional\tabularnewline
$\mathcal{L}$ & Lagrangian\tabularnewline
$\tilde{n}$ & Target density\tabularnewline
$N_{\text{orbs}}$ & Number of Kohn-Sham orbitals\tabularnewline
$\phi^{m}$ and $\chi^{m}$ & m\textsuperscript{th} orbital and corresponding adjoint orbital\tabularnewline
$\underline{\phi}$ & All $N_{\text{orbs}}$ orbitals grouped into a vector\tabularnewline
$w_{{\bf i}}$ & Optional cost-functional weight\tabularnewline
$dx_{{\bf i}}$ and $dk_{{\bf j}}$ & Integration weights\tabularnewline
\end{tabular}
\end{table}


The notation used in the following derivations is the same notation used in Ref.~\citenum{jensen_numerical_2016} for \gls{tddft} inverse problems.
It is very similar to the continuous adjoint derivation found in quantum optimal control derivations applied to \gls{tddft}.\cite{castro_controlling_2012}
The discrete adjoint method is discussed in depth in Ref.~\citenum{giles_introduction_2000}.
The notation used here is simply a modification of other discrete adjoint derivations applied to the \gls{dft} inverse problem.

The discretized cost functional
\begin{align}
F\left[v\right] & =F\left[\underline{\phi}\left[v\right]\right]=\frac{1}{2}\sum_{{\bf i}={\bf 1}}^{N_{{\bf x}}}\left(\sum_{m=1}^{N_{\text{orbs}}}\phi_{{\bf i}}^{m\,*}\phi_{{\bf i}}^{m}-\tilde{n}_{{\bf i}}\right)^{2}w_{{\bf i}}\label{eq:costfunc-orbs-ti}
\end{align}
can be differentiated with respect to an arbitrary orbital $\phi_{\boldsymbol{\alpha}}^{\gamma}$ to produce the expression
\begin{align}
\frac{\partial F}{\partial\phi_{\boldsymbol{\alpha}}^{\gamma}} & =w_{\boldsymbol{\alpha}}\left(\sum_{m=1}^{N_{\text{orbs}}}\phi_{\boldsymbol{\alpha}}^{m\,*}\phi_{\boldsymbol{\alpha}}^{m}-\tilde{n}_{{\bf i}}\right)\phi_{\boldsymbol{\alpha}}^{\gamma\,*},\label{eq:costfunc-deriv-ti}
\end{align}
where $\gamma$ is the orbital index and $\boldsymbol{\alpha}=\alpha_{x},\alpha_{y},\alpha_{z}$ is the spatial index of the orbital being differentiated.
We introduce the Lagrangian
\begin{align}
\mathcal{L}\left[v,\underline{\phi},\underline{\chi}\right] & =2\Re\left\{ \sum_{{\bf i}={\bf 1}}^{N_{{\bf x}}}\sum_{m=1}^{N_{\text{orbs}}}\chi_{{\bf i}}^{m}\left[H\left(v\right)\phi^{m}-\varepsilon^{m}\phi^{m}\right]_{{\bf i}}^{*}\right\} +\sum_{m=1}^{N_{\text{orbs}}}\lambda^{m}\left[\left(\phi_{{\bf i}}^{m}\right)^{*}\phi_{{\bf i}}^{m}-\tilde{n}_{{\bf i}}\right]dx_{{\bf i}}.\label{eq:lagrangian-ti}
\end{align}
where the Lagrange multipliers $\left\{ \chi^{m}\right\} $ and $\left\{ \lambda^{m}\right\} $ constrain the orbitals to satisfy Eqs.~\eqref{eq:ks-eval-inverse} and \eqref{eq:dft-orthonormal}.
The Hamiltonian $H\left(v\right)=-\frac{\Delta}{2}+v$ in Eq.~\eqref{eq:lagrangian-ti} is 
is a discretized version of Eq.~\eqref{eq:ks-eval-inverse}.
We subtract the Lagrangian from the cost functional to form the total functional
\begin{equation}
J\left[v,\underline{\phi},\underline{\chi}\right]=F\left[\underline{\phi}\right]-\mathcal{L}\left[v,\underline{\phi},\underline{\chi}\right].\label{eq:total-functional-ti}
\end{equation}

Differentiating the total functional $J$ with respect to the adjoint orbitals $\left\{ \chi^{m}\right\} $  and normalization multipliers $\left\{ \lambda^{m}\right\} $ gives the standard \gls{ks} equations
\begin{align}
0=\frac{\partial J}{\partial\chi_{\boldsymbol{\alpha}}^{\gamma}}\,\forall\boldsymbol{\alpha},\gamma & \,\Rightarrow\,H\!\left(v\right)\phi^{m}=\varepsilon^{m}\phi^{m}\text{ and}\label{eq:forward-ti}\\
0=\frac{\partial J}{\partial\varepsilon^{\gamma}}\,\forall\gamma & \,\Rightarrow\,\sum_{{\bf i}={\bf 1}}^{N_{{\bf x}}}\left(\phi_{{\bf i}}^{m}\right)^{*}\phi_{{\bf i}}^{m}\cdot dx_{{\bf i}}=1.\label{eq:normalization-ti}
\end{align}
The adjoint equations are found by differentiating with respect to the \gls{ks} orbitals and eigenvalues to produce a set of sparse linear equations
\begin{equation}
\left[\begin{array}{c}
\left(\frac{\partial F}{\partial\phi^{\gamma}}\right)^{*}\\
\left(\frac{\partial F}{\partial\varepsilon^{\gamma}}\right)^{*}
\end{array}\right]=\left[\begin{array}{cc}
H^{\dagger}\!\left(v\right)-\left(\varepsilon^{\gamma}\right)^{*} & \phi^{\gamma}\\
-\left(\phi^{\gamma}\right)^{\dagger} & 0
\end{array}\right]\left[\begin{array}{c}
\chi^{\gamma}\\
\lambda^{\gamma}
\end{array}\right],\label{eq:sparse-linear-ti}
\end{equation}
where $\frac{\partial F}{\partial\varepsilon^{\gamma}}=0$.
After solving the direct and adjoint equations in Eqs.~\eqref{eq:forward-ti}-\eqref{eq:sparse-linear-ti}, the total derivative is given by the expression
\begin{equation}
\frac{\partial J}{\partial v_{\boldsymbol{\alpha}}}=-\frac{\partial\mathcal{L}}{\partial v_{\boldsymbol{\alpha}}}=-2\sum_{m=1}^{N_{\text{orbs}}}\sum_{{\bf i}={\bf 1}}^{N_{{\bf x}}}\Re\left\{ \left(\chi_{{\bf i}}^{m}\right)^{*}\frac{\partial H\!\left(v\right)}{\partial v_{\boldsymbol{\alpha}}}\phi_{{\bf i}}^{m}\right\} .\label{eq:total-deriv-ti}
\end{equation}
If the potential is placed diagonally along $H$ then this last expression reduces to 
\begin{equation}
\frac{\partial J}{\partial v_{\boldsymbol{\alpha}}}=-2\sum_{m=1}^{N_{\text{orbs}}}\Re\left\{ \left(\chi_{\boldsymbol{\alpha}}^{m}\right)^{*}\phi_{\boldsymbol{\alpha}}^{m}\right\} .\label{eq:total-deriv-ti-simplified}
\end{equation}

A very similar derivation is found when using the scaled orbitals described in \myref{sec:pde-constrained}.
In this case, the discretized cost functional is
\begin{align}
F\left[v\right] & =F\left[\underline{g}\left[v\right]\right]=\frac{1}{2}\sum_{{\bf i}={\bf 1}}^{N_{{\bf x}}}\left(\sum_{m=1}^{N_{\text{orbs}}}\left|g_{{\bf i}}^{m}\right|^{2}-1\right)^{2}w_{{\bf i}},\label{eq:costfunc-scaled}
\end{align}
where $\phi_{i}^{m}=\sqrt{\tilde{n}_{{\bf i}}}g_{{\bf i}}^{m}$ and 
\begin{align}
\frac{\partial F}{\partial g_{\boldsymbol{\alpha}}^{\gamma}} & =\left(\sum_{m=1}^{N_{\text{orbs}}}\left|g_{\boldsymbol{\alpha}}^{m}\right|^{2}-1\right)\left(g_{\boldsymbol{\alpha}}^{\gamma}\right)^{*}w_{\boldsymbol{\alpha}}.\label{eq:costfunc-deriv-scaled}
\end{align}
The forward equations are
\begin{align}
H\!\left(v\right)g^{m}=\varepsilon^{m}g^{m}\text{ and}\label{eq:forward-scaled}\\
\sum_{{\bf i}={\bf 1}}^{N_{{\bf x}}}\tilde{n}_{{\bf i}}g_{{\bf i}}^{m\,*}g_{{\bf i}}^{m}dx_{{\bf i}}=1.\label{eq:normalization-scaled}
\end{align}
The corresponding adjoint equation is
\begin{equation}
\left[\begin{array}{c}
\left(\frac{\partial F}{\partial g^{\gamma}}\right)^{*}\\
\left(\frac{\partial F}{\partial\varepsilon^{\gamma}}\right)^{*}
\end{array}\right]=\left[\begin{array}{cc}
H^{\dagger}\!\left(v\right)-\left(\varepsilon^{\gamma}\right)^{*} & \tilde{n}g^{\gamma}dx\\
-\left(g^{\gamma}\right)^{\dagger} & 0
\end{array}\right]\left[\begin{array}{c}
\chi^{\gamma}\\
\lambda^{\gamma}
\end{array}\right]\label{eq:sparse-linear-scaled}
\end{equation}
and the total derivative is
\begin{equation}
\frac{\partial J}{\partial v_{\boldsymbol{\alpha}}}=-2\sum_{m=1}^{N_{\text{orbs}}}\sum_{{\bf i}={\bf 1}}^{N_{{\bf x}}}\Re\left\{ \left(\chi_{{\bf i}}^{m}\right)^{*}\frac{\partial H\!\left(v\right)}{\partial v_{\boldsymbol{\alpha}}}g_{{\bf i}}^{m}\right\} ,\label{eq:total-deriv-scaled}
\end{equation}
where $H$ is a discretization of the scaled Hamiltonian given in Eq.~\eqref{eq:ks-eval-expanded}.

The discrete adjoint equations for a periodic system follow the same pattern as the above derivations but with the added complication of a numerical integration in $k$-space.
The discretized cost functional is
\begin{align}
F\left[v\right] & =F\left[\underline{\phi}\left[v\right]\right]=\frac{1}{2}\sum_{{\bf i}={\bf 1}}^{N_{{\bf x}}}w_{{\bf i}}\left(\sum_{m=1}^{N_{\text{orbs}}}\sum_{{\bf j}={\bf 1}}^{N_{{\bf k}}}u_{{\bf j}}\left|\phi_{{\bf i},{\bf j}}^{m}\right|^{2}-\tilde{n}_{{\bf i}}\right)^{2},\label{eq:costfunc-periodic}
\end{align}
where $u_{{\bf j}}=\frac{1}{\Omega_{{\bf k}}}dk_{{\bf j}}$ is the
$j^{\text{th}}$ integration weight in $k$-space divided by the volume
of the first Brillouin zone and $\phi_{{\bf i},{\bf j}}^{m}$ corresponds
to the point $\phi_{{\bf k}_{{\bf j}}}^{m}\left({\bf x}_{{\bf i}}\right)$.
The forward equations are 
\begin{align}
H_{{\bf j}}\!\left(v\right)\phi_{,{\bf j}}^{m}=\varepsilon_{{\bf j}}^{m}\phi_{,{\bf j}}^{m}\text{ and}\label{eq:forward-periodic}\\
\sum_{{\bf i}={\bf 1}}^{N_{{\bf x}}}\left(\phi_{{\bf i},{\bf j}}^{m}\right)^{*}\phi_{{\bf i},{\bf j}}^{m}dx_{{\bf i}}=1,\label{eq:normalization-periodic}
\end{align}
where $H_{{\bf j}}\!\left(v\right)$ is a discretization of the Hamiltonian
$-\frac{1}{2}\left(\Delta+2\imath{\bf k}_{{\bf j}}\cdot\boldsymbol{\nabla}-k_{{\bf j}}^{2}\right)+v\!\left({\bf r}\right)$.
The corresponding adjoint equations are
\begin{equation}
\left[\begin{array}{c}
\left(\frac{\partial F}{\partial\phi_{\boldsymbol{\alpha},\boldsymbol{\beta}}^{\gamma}}\right)^{*}\\
\left(\frac{\partial F}{\partial\varepsilon_{\boldsymbol{\beta}}^{\gamma}}\right)^{*}
\end{array}\right]=\left[\begin{array}{cc}
\left(H_{\boldsymbol{\beta}}\right)^{\dagger}\!\left(v\right)-\left(\varepsilon_{\boldsymbol{\beta}}^{\gamma}\right)^{*} & \phi_{,\boldsymbol{\beta}}^{\gamma}dx\\
-\left(\phi_{,\boldsymbol{\beta}}^{\gamma}\right)^{\dagger} & 0
\end{array}\right]\left[\begin{array}{c}
\chi_{,\boldsymbol{\beta}}^{\gamma}\\
\lambda_{\boldsymbol{\beta}}^{\gamma}
\end{array}\right]\label{eq:sparse-linear-periodic}
\end{equation}
and the total derivative is
\begin{equation}
\frac{\partial J}{\partial v_{\boldsymbol{\alpha}}}=-\frac{\partial\mathcal{L}}{\partial v_{\boldsymbol{\alpha}}}=-2\sum_{{\bf j}={\bf 1}}^{N_{{\bf k}}}\sum_{m=1}^{N_{\text{orbs}}}\sum_{{\bf i}={\bf 1}}^{N_{{\bf x}}}\Re\left\{ \left(\chi_{{\bf i},{\bf j}}^{m}\right)^{*}\frac{\partial H_{{\bf j}}\!\left(v\right)}{\partial v_{\boldsymbol{\alpha}}}\phi_{{\bf i},{\bf j}}^{m}\right\} .\label{eq:total-deriv-periodic}
\end{equation}

\end{appendices}

\clearpage


\bibliography{refs}

\end{document}